\documentclass{article}

\usepackage{arxiv}

\pdfoutput=1

\usepackage{graphicx}
\usepackage[usenames,dvipsnames]{xcolor} % colour
\usepackage{amssymb,mathtools,bm}
\usepackage{amsfonts} %my add
\usepackage{epsfig,subfigure} % plots
\usepackage{booktabs,longtable} % tables
\usepackage{exscale,relsize} % scale 
\usepackage[normalem]{ulem} % editing
\usepackage{enumerate} 
\usepackage{hyperref}
\usepackage{float}
\usepackage[utf8]{inputenc}
\usepackage{comment}
\usepackage{color,colortbl}
\usepackage{tensor}

\usepackage{multirow,array}

\usepackage{adjustbox}  %resize table size

\newcommand{\w}[1]{\bm{#1}}

\newcommand{\M}{\mathcal{M}}

\newcommand{\T}{\mathcal{T}}

\newcommand{\tc}{{\tilde c}}

\newcommand{\of}[1]{\underset{\sim}{#1}} %My add one-form =\of
\newcommand{\tf}[1]{\underset{\approx}{#1}} %My add two-form = \tf
\newcommand{\ovec}[1]{\overline{#1}} %My add vector =\ovec
\newcommand{\tvec}[1]{\overset{=}{#1}} %My add Rancktwo tensor= \tvec

\title{3+1 formalism of the minimally extended varying speed of light model}

\author{
 Seokcheon Lee \\
  Department of Physics, Institute of Basic Science\\
  Sungkyunkwan University, \\
  Suwon 16419, Korea \\
  \texttt{skylee@skku.edu} \\
}

\begin{document}
\maketitle
\begin{abstract}
The $3+1$ formalism provides a structured approach to analyzing spacetime by separating it into spatial and temporal components. When applied to the Robertson-Walker metric, it simplifies the analysis of cosmological evolution by dividing the Einstein field equations into constraint and evolution equations. It introduces the lapse function $N$ and the shift vector $N^i$, which control how time and spatial coordinates evolve between hypersurfaces. In standard model cosmology, $N = 1$ and $N^i = 0$ for the Robertson-Walker metric. However, the $N$ becomes a function of time when we apply the metric to the minimally extended varying speed of light model. This approach allows for a more direct examination of the evolution of spatial geometry and offers flexibility in handling scenarios where the lapse function and shift vector vary. In this manuscript, we derive the model's $N$, $N^i$, along with the constraint and evolution equations, and demonstrate their consistency with the existing Einstein equations. We have shown in a previous paper that the possibility of changes in the speed of light in the Robertson-Walker metric is due to cosmological time dilation. Through the $3+1$ formalism, we can make the physical significance more explicit and demonstrate that it can be interpreted as the lapse function. From this, we show that the minimally extended varying speed of light model is consistent.
\end{abstract}

% keywords can be removed

\section{Introduction}
\label{sec:intro}

In Einstein's field equations (EFEs), space and time are treated equally, reflecting their theoretical covariance. However, this symmetry can make it challenging to understand the temporal evolution of the gravitational field. Therefore, when solving EFEs to study this evolution, the first step is to separate the roles of space and time. The Arnowitt-Deser-Misner (ADM) formalism is a potent approach in general relativity (GR) used to describe the dynamics of spacetime by decomposing it into a family of spacelike hypersurfaces parameterized by time \cite{Arnowitt:1962hi,Corichi:1991qqo}. A foliation of spacetime into spacelike hypersurfaces, indexed by a time parameter $t$, represents this decomposition. It is a specific application of the $3+1$ formalism. It expresses EFEs in Hamiltonian form, introducing canonical variables (like the spatial metric and its conjugate momentum) and deriving the Hamiltonian and momentum constraints. Since the mid-1980s, interest in ADM formalism has grown, both in formalism and its applications. Notable examples include its role in developing Ashtekar variables \cite{Ashtekar:1987prd}, its use in gravity quantization via path integrals \cite{Guven:1991eq}, its application in numerically solving EFEs \cite{Hilditch:2024nhf}, and its contribution to the theory of the space of spaces \cite{DeWitt:1967yk}.

The $3+1$ formalism is an approach in GR that reformulates the EFEs by splitting spacetime into three spatial dimensions and one temporal dimension \cite{Gourgoulhon:2007ue,Alcubierre:08,Baumgarte:10,Shibata:15,Palenzuela:2020tga}. The $3+1$ formalism is a broader concept, referring to any decomposition of spacetime into space and time components. The ADM formalism is a specific application of it focusing on the Hamiltonian structure of GR. It allows the complete division of a globally hyperbolic spacetime into three-dimensional spatial slices. These slices can be viewed as level sets of a parameter $t$, which acts as a universal time function. It is important to note that $t$ does not necessarily match the proper time of any particular observer. This division of spacetime into spatial hypersurfaces is often referred to as synchronization. In the Robertson-Walker (RW) metric, the universal time function $t$ does coincide with the proper time of observers moving with the expansion of the universe \cite{Islam01,Narlikar02,Hobson06,Roos15}. %Introduction to 3+1 Numerical Relativity (M. Alcubierre) p.80

The $3+1$ formalism provides significant advantages for solving the EFEs, especially in contexts where separating space and time is crucial~\cite{Gourgoulhon:2007ue,Alcubierre:08,Baumgarte:10,Shibata:15,Palenzuela:2020tga}. It frames GR as an initial value problem, enabling the evolution of spatial geometries over time, which is essential for simulations like black hole (BH) mergers or gravitational wave (GW) studies. It is precious in cosmology, as it splits spacetime into spatial hypersurfaces that evolve, allowing for studying how spatial geometry changes. In this formalism, we decompose the spacetime metric into the spatial metric, the lapse function, and the shift vector. This approach separates the EFEs into constraint and evolution equations, simplifying the problem into manageable parts \cite{Larena:2009md,Chaichian:2010yi,Carloni:2010nx,Ardehali:2017rgd,Li:2018hll,Bahamonde:2021gfp,Cabass:2022avo,Gaur:2022hap}. It also provides flexibility in choosing coordinates, such as using comoving coordinates in cosmology. It is compatible with gauge theories, enhancing the understanding and computation of gauge freedoms in cosmological perturbation theory \cite{Alles:2015vua,Gong:2017tev,Magi:2022nfy,Chataignier:2023rkq}.

When we observe distant astronomical objects, such as type Ia supernovae (SNeIa) \cite{Leibundgut:1996qm,SupernovaSearchTeam:1997gem,Foley:2005qu,Blondin:2007ua,Blondin:2008mz,DES:2024vgg}, gamma-ray bursts (GRBs)  \cite{Norris:1993hda,Wijers:1994qf,Band:1994ee,Meszaros:1995gj,Lee:1996zu,Chang:2001fy,Crawford:2009be,Zhang:2013yna,Singh:2021jgr}, or quasars (QSOs) \cite{Hawkins:2001be,Dai:2012wp,Lewis:2023jab}, the light from these objects has traveled through an expanding universe. As a result, the time interval between the arrival of successive light pulses (or any periodic signal) appears stretched due to the expansion. This stretching of time intervals is what we refer to as cosmological time dilation (CTD). If an event occurs at a time $t_{\text{em}}$ (emission time) in the past when the scale factor was $a(t_{\text{em}})$, and we observe it at present $t_0$, the time interval $\Delta t_{\text{obs}}$ between observed successive pulses will be longer than the interval $\Delta t_{\text{em}}$ at the time of emission. The relationship is given by
\begin{align}
\Delta t_{\text{obs}} = \frac{1}{a(t_{\text{em}})^{1+\beta}} \Delta t_{\text{em}} = \left( 1 + z \right)^{1+\beta} \label{Deltat} \,,
\end{align}
where we use $a(t_0) = 1$ and $1 + z = \frac{1}{a(t_{\text{em}})}.$ Therefore, we can connect CTD to the redshift $z$ of the object. The higher the redshift (i.e., the further back in time we are looking), the more significant the time dilation effect. This CTD means that if a supernova, for example, had a light curve (LC) with a specific duration when it exploded, that duration would appear longer to us by the factor $(1+z)^{1+\beta}$ due to the universe's expansion. In the standard model of cosmology (SMC), we assume that $\beta = 0$. However, the RW metric does not require this. Thus, if observations find any evidence of non-zero $\beta$, then there is a possibility that the speed of light may vary with cosmic time \cite{Lee:2020zts,Lee:2022heb,Lee:2023bjz,Lee:2024part,Lee:2024des,Bileska:2024odt}. We call it the minimally extended varying speed of light (meVSL) with $c = c_0 a^{b/4}$. In this model, $b = -4\beta$. We show constraints on the $\beta$ obtained from each observation in Table~\ref{tab:CTDs} \cite{Lee:2024des}. The Dark Energy Survey (DES) analyzed SNeIa LCs in multiple bands, finding that the CTD effect scales as $(1+z)^{1.003\pm0.005}$ across all bands, supporting the CTD effect with consistent results. An analysis of $247$ GRBs indicated that the power-law index $B$ aligns with a cosmological signature within $1$-$\sigma$, supporting the presence of CTD. For QSOs, variability was analyzed using LCs, finding a CTD-related factor of $1.28^{+0.28}_{-0.29}$, consistent with the expected effect. 

 \begin{table}[h!]
 	\begin{center}
	%\begin{adjustbox}{width=\columnwidth,center}
		\begin{tabular}{|c|c|c|c|} 
			\hline
			obs  & $1+\beta$ & $\#$ of samples & ref \\ \hline	
			\multirow{2}{*}{SNeIa} \quad i-band  & $0.988\pm0.008$  & 1465 & \multirow{2}{*}{\cite{DES:2024vgg}} \\ 
			\qquad \quad \,\, $4$-bands  & $1.003\pm0.005$ & 1504 &  \\ \hline
			\multirow{4}{*}{GRBs} \multirow{2}{*}{unbinned} \,\, $T_{50}$& $0.66^{+0.17}_{-0.17}$ & \multirow{4}{*}{$247$} &  \multirow{4}{*}{\cite{Singh:2021jgr}}  \\
				\hspace{3.0cm} $T_{90}$ & $0.52^{+0.15}_{-0.16}$ & & \\ 
			\qquad \quad	\multirow{2}{*}{binned} \quad \, $T_{50}$ &$1.18^{+0.26}_{-0.36}$  & & \\ 
				\hspace{3.0cm} $T_{90}$  & $0.97^{+0.29}_{-0.30}$ & & \\ \hline	
			QSOs & $1.28^{+0.28}_{-0.29}$ &190 & \cite{Lewis:2023jab} \\ \hline			
\end{tabular}
%\end{adjustbox}
\end{center}
\caption{This table summarizes the most recent CTD data obtained from SNeIa, GRBs, and QSOs.}
\label{tab:CTDs}
 \end{table}

In the varying speed of light model, the speed of light has a constant value on a single hypersurface to preserve the local Lorentz invariant, but as this evolves, it can change. Therefore, the speed of light on a hypersurface at a different time will vary, allowing for a clear description of the energy-momentum stress tensor accordingly \cite{Lee:2020zts,Lee:2022heb,Lee:2023bjz,Lee:2024part,Lee:2024des}.

When we apply the $3+1$ formalism to the RW metric, the shift vector is zero because the metric's homogeneity and isotropy imply no preferred spatial direction. Consequently, the coordinates are chosen so that observers move orthogonally to the spatial hypersurfaces, resulting in no spatial coordinate shifts over time.  In the RW metric, proper time and physical time are identical for comoving observers, who move with the cosmic expansion. Proper time experienced by an observer along their worldline matches the coordinate time $t$ in the RW metric \cite{Islam01}. Since the metric describes a homogeneous and isotropic universe, all comoving observers share this same proper time, making it equivalent to physical time in this context. However, for non-comoving observers, proper time may differ from coordinate time.  Thus,  the lapse function is $1$ in the RW metric. This conclusion holds in SMC. However, in the meVSL model, the lapse function may be defined differently, and in this case, it could be a function of time.

In this manuscript, we demonstrate that using the $3+1$ formalism allows for a clear connection between variations in the speed of light and the lapse function within the meVSL model. We also show that by employing constraint and evolution equations at each time point (\textit{i.e.}, each hypersurface), variations of the speed of light are correctly applied to the full EFEs \cite{Lee:2020zts,Lee:2022heb,Lee:2023bjz,Lee:2024part}. This approach enhances the physical understanding of the meVSL model. Furthermore, we expect it to help resolve gauge choice issues in the perturbation equations of the field equations, which we will address in detail in a forthcoming paper.

In the following section, we briefly review the meVSL model. In section \ref{sec:ADM}, we provide a detailed explanation of the $3+1$ formalism using the RW metric within the meVSL model. First, we describe Eulerian observers based on the RW metric in the meVSL framework. We also discuss intrinsic curvature, extrinsic curvature, and the Gauss-Codazzi relation. Later, we use these concepts to find the constraint and evolution equations. We do this by projecting the stress-energy tensor and Einstein tensors onto the hypersurface and along the normal direction in the EFEs. We discuss it in Section \ref{sec:EEs}. In Section \ref{sec:con}, we provide a discussion and conclusion on this method.

\section{Brief Review of the Minimally Extended Varying Speed-of-Light (meVSL) Model}
\label{Sec:meVSL} 

In this session, we review the potential of the VSL model and introduce the meVSL model~\cite{Lee:2020zts, Lee:2022heb, Lee:2023bjz, Lee:2024part, Lee:2024des, Lee:2023rqv, Lee:2023ucu}. By adopting the cosmological principle with incorporating Weyl’s postulate, we yield the line element 
\begin{align}
ds^2 = - c(t)^2 dt^2 + a(t)^2 \left[ \frac{dr^2}{1-Kr^2} + r^2  \left( d \theta^2 + \sin^2 \theta d \phi^2 \right)  \right] \equiv - c(t)^2 dt^2 + a(t)^2 dl_{3\textrm{D}}^2 \label{dstgen} \,. 
\end{align}

\begin{figure}[H]
 \includegraphics[width=0.9\textwidth]{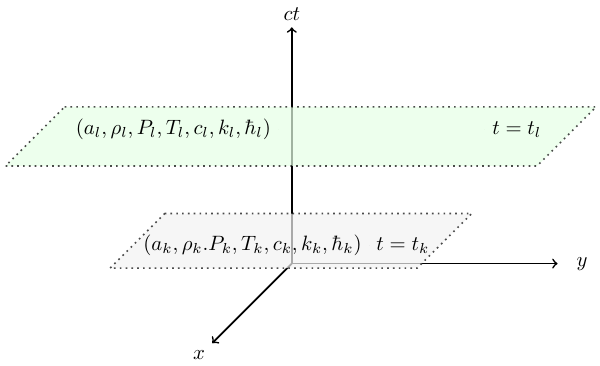} 
	\caption{At \( t = t_k \), physical quantities and constants, such as \( a_k \), \( \rho_k \), \( P_k \), \( T_k \), \( c_k \), \( k_k \), and \( \hbar_k \), are fixed and uniform across the spatial hypersurface defined by \( t = t_k \). As the universe evolves and expands, these quantities change to \( a_l \), \( \rho_l \), \( P_l \), \( T_l \), \( c_l \), \( k_l \), and \( \hbar_l \) at time \( t = t_l \). Importantly, the CP and Weyl’s postulate do not require the speed of light \( c_k \) at time \( t_k \) to be equal to \( c_l \) at time \( t_l \); instead, its value is governed by the cosmological time dilation relation.}
	\label{Fig1}
 
\end{figure}

Here, the speed of light is a function of time, differing from the conventional RW metric. While the RW metric assumes quantities like the scale factor \(a_l = a(t_l)\), mass density \(\rho_l = \rho(t_l)\), pressure \(P_l = P(t_l)\), and others remain constant on hypersurfaces defined by constant \(t_l\) or \(t_k\), Weyl’s postulate allows these to vary with cosmic time \( t \), accounting for cosmological redshift (see Figure~\ref{Fig1}). Traditionally, physical constants, including the speed of light, are assumed constant over cosmic time. However, this assumption that \( c_l = c_k \) is not directly related to the conditions required to derive the RW metric, and it relies on CTD. GR does not specify laws governing this constancy. As the Universe evolves from \( t_k \) to \( t_l \), quantities like \(a(t)\), \(\rho(t)\), \(P(t)\), and \(T(t)\) change with time, determined by the solution of Einstein's Field Equations (EFEs) and Bianchi’s identity (BI), along with the equation of state~\cite{Lee:2020zts, Lee:2023bjz}. 

\vspace{-6pt}

\subsection{Cosmological Redshift}
\label{subsec:CRedshift}

The RW metric in Equation~\eqref{dstgen} can be rewritten as
\begin{align}
ds^2 = -(d X^0)^2 + a^2(t) \left( \frac{dr^2}{1-kr^2} + r^2 \left( d \theta^2 + \sin^2 \theta d \phi^2 \right) \right) \equiv -(d X^0)^2 + a^2(t) dl_{3\textrm{D}}^2 \label{RW} \,,
\end{align}
where \( X^0 = c t \). The light signal propagates along the null geodesic \( ds^2 = 0 \), yielding the outgoing light signals
\begin{align}
dl_{3\textrm{D}}(r, \theta, \phi) = \frac{dX^0}{a(t)} \label{dl3D} \,.
\end{align}

The spatial infinitesimal line element \( dl_{3\textrm{D}} \) is a function of comoving coordinates only, implying that it is constant at any given time. From this, the standard model of cosmology (SMC), assuming a constant speed of light, derives the cosmological redshift relation
\begin{align}
\frac{dX^0(t_1)}{a(t_1)} = \frac{dX^0(t_2)}{a(t_2)} \quad \Rightarrow \quad \frac{dt_1}{a_1} = \frac{dt_2}{a_2} \quad \Rightarrow \quad \lambda_1 = \frac{a_1}{a_2} \lambda_2 \label{z} \,.
\end{align}

This redshift is traditionally derived assuming a constant speed of light. However, since Lorentz invariance (LI) is a local symmetry and GR holds at cosmological scales, the validity of special relativity (SR) at cosmological distances should be observationally determined~\cite{Roos15}. Allowing the speed of light to vary over time, we rewrite Equation~\eqref{z} as
\begin{align}
\frac{c_1 dt_1}{a_1} = \frac{c_2 dt_2}{a_2} \quad \Rightarrow \quad \lambda_1 = \frac{a_1}{a_2} \lambda_2 \label{zVSL} \,,
\end{align}
where \( c_i \equiv c(t_i) \) and \( a_i \equiv a(t_i) \). We express
\begin{align}
dX^{0} = d \left(\tilde{c} t \right) = \left(\frac{d \ln \tilde{c}}{d \ln t} + 1 \right) \tilde{c} dt \equiv c dt \quad \textrm{and} \quad \delta c \equiv \frac{c}{\tilde{c}} = \left(\frac{d \ln \tilde{c}}{d \ln t} + 1 \right) \label{dX0} \,.
\end{align}

Thus, the cosmological redshift relation holds even when the speed of light varies as a function of cosmic time, as derived in the meVSL model~\cite{Lee:2020zts}.

\subsection{The Possibility of Varying Speed-of-Light Theory in the Robertson--Walker Metric}
\label{subsec:FLRW}

The redshift derivation involves the geodesic equation for light, where \( ds^2 = 0 \) as in Equation~\eqref{dstgen}. The consistency of \( dl_{3\textrm{D}} \) over time is ensured by using comoving coordinates. Expanding on this, the outgoing light signals are expressed as
\begin{align}
d l_{3\textrm{D}} &= \frac{c(t_i) dt_i}{a(t_i)} \quad : \quad \frac{c_1 dt_1}{a_1} = \frac{c_2 dt_2}{a_2} \Rightarrow \begin{cases} 
c_1 = c_2 = c & \textrm{if} \quad \frac{dt_1}{a_1} = \frac{dt_2}{a_2} \quad \textrm{SMC} \\ 
c_1 = \frac{f(a_2)}{f(a_1)} \frac{a_1}{a_2} c_2 & \textrm{if} \quad \frac{dt_1}{f(a_1)} = \frac{dt_2}{f(a_2)} \quad \textrm{VSL} \\ 
c_1 = \left( \frac{a_1}{a_2}\right)^{\frac{b}{4}} c_2 & \textrm{if} \quad \frac{dt_1}{a_1^{1-\frac{b}{4}}} = \frac{dt_2}{a_2^{1-\frac{b}{4}}} \quad \textrm{meVSL}  
\end{cases} \,, \label{dl3D2}
\end{align}
where \( dt_i = 1/\nu(t_i) \) is the time interval between light crests, and \( f(a_i) \) is an arbitrary function of \( a(t_i) \).

In the SMC, it is assumed that the cosmological time dilation (TD) between two hypersurfaces at \( t_1 \) and \( t_2 \) is proportional to the inverse of the scale factors \( a(t) \), implying constant speed of light \( c(t_1) = c(t_2) = c \). However, this assumption lacks physical justification in GR, where the constancy of light speed holds only locally. In an expanding universe, the scale factor increases, leading to cosmological redshift of various quantities. However, cosmological TD cannot be derived from the CP and Weyl’s postulate alone. It must be determined experimentally. 

\subsection{The Modification of Einstein's Field Equations}
\label{subsec:EFE}

In the meVSL model, the line element is given by
\begin{align}
ds^2 = - c^2 dt^2 + a^2 \gamma_{\ij} dx^{i} dx^{j} \label{dsFRW} \,.
\end{align}
The Riemann curvature tensors, Ricci tensors, and Ricci scalar are expressed as:
\begin{align}
\tensor{R}{^0_i_0_j} &= \frac{g_{ij}}{c^2} \left( \frac{\ddot{a}}{a} - H^2 \frac{d \ln c}{d \ln a} \right) , \quad
\tensor{R}{^i_0_0_j} = \frac{\delta^{i}_{j}}{c^2} \left( \frac{\ddot{a}}{a} - H^2 \frac{d \ln c}{d \ln a} \right) , \nonumber \\
\tensor{R}{^i_j_k_m} &= \left( \frac{H^2}{c^2} + \frac{k}{a^2} \right) \left( \delta^{i}_{k} g_{jm} - \delta^{i}_{m} g_{jk} \right) \label{tRijkmApp} \,, \\
R_{00} &= - \frac{3}{c^2} \left( \frac{\ddot{a}}{a} - H^2 \frac{d \ln c}{d \ln a} \right)  \quad , \quad
R_{ii} = \frac{g_{ii}}{c^2} \left( 2 \frac{\dot{a}^2}{a^2} + \frac{\ddot{a}}{a} + 2 k \frac{c^2}{a^2} - H^2 \frac{d \ln c}{d \ln a} \right) \,, \nonumber \\
R &= \frac{6}{c^2} \left( \frac{\ddot{a}}{a} + \frac{\dot{a}^2}{a^2} + k \frac{c^2}{a^2} - H^2 \frac{d \ln c}{d \ln a} \right) \label{tRmp} \,.
\end{align}
The energy-momentum tensor is
\begin{align}
T_{\mu}^{\nu} = \text{diag} \left(-\rho c^2, P, P, P \right) \label{Tmunump} \,.
\end{align}
The energy conservation is given by Bianchi's identity:
\begin{align}
\rho_i c^2 = \rho_{i0} c_0^2 a^{-3 (1 + \omega_i)} \label{rhotc2mp} \,,
\end{align}
where \( c_0 \) and \( \rho_{i0} \) are the present values of the speed of light and mass density, respectively, and \( a_0 = 1 \).

The EFEs, including the cosmological constant, are:
\begin{align}
\frac{\dot{a}^2}{a^2} + k \frac{c^2}{a^2} - \frac{\Lambda c^2}{3} &= \frac{8 \pi G}{3} \sum_{i} \rho_i \label{tG00mp} \,, \\
\frac{\dot{a}^2}{a^2} + 2 \frac{\ddot{a}}{a} + k \frac{c^2}{a^2} - \Lambda c^2 - 2 H^2 \frac{d \ln c}{d \ln a} &= -8 \pi G \sum_{i} \frac{P_i}{c^2} = -8 \pi G \sum_{i} \omega_{i} \rho_{i} \label{tG11mp} \,.
\end{align}
Subtracting Equation~\eqref{tG00mp} from Equation~\eqref{tG11mp} gives:
\begin{align}
\frac{\ddot{a}}{a} = -\frac{4 \pi G}{3} \sum_{i} \left( 1 + 3 \omega_i \right) \rho_i + \frac{\Lambda c^2}{3} + H^2 \frac{d \ln c}{d \ln a} \label{tG11mG00mp} \,.
\end{align}
This equation shows that the expansion and acceleration of the Universe depend on \( c \), \( G \), and \( \rho \). By differentiating Equation~\eqref{tG00mp} with respect to cosmic time and using Equation~\eqref{rhotc2mp}, we obtain the relation between \( G \) and \( c \):
\begin{align}
\frac{d \ln G}{d \ln a} = 4 \frac{d \ln c}{d \ln a} \equiv b = \text{const.} \quad \Rightarrow \quad \frac{G}{G_0} = \left( \frac{c}{c_0} \right)^4 = \left( \frac{a}{a_0} \right)^{b} \label{dotGoGmp} \,.
\end{align}
From this, the time variations of \( c \) and \( G \) are:
\begin{align}
\frac{\dot{G}}{G} = b H \quad , \quad \frac{\dot{c}}{c} = \frac{b}{4} H \label{dotcoc} \,.
\end{align}
Thus, the time variation ratios for \( G \) and \( c \) are:
\begin{align}
\frac{\dot{G}_0}{G_0} = b H_0 \quad , \quad \frac{\dot{c}_0}{c_0} = \frac{b}{4} H_0 \label{dotcoc0} \,.
\end{align}

For more detailed information, please refer to Reference~\cite{Lee:2020zts}.
In the meVSL model, local thermodynamics, energy conservation, and other physical laws dictate the time evolution of various physical constants and quantities. The relationships governing these evolutions are summarized in Table~\ref{tab:table-1}.

\begin{table}[htbp]
	%\centering
\caption{Summary of the cosmological evolution of physical constants and quantities in the meVSL model. These relationships are consistent with all known local physical laws, including special relativity, thermodynamics, and electromagnetism \cite{Lee:2020zts}.}
\label{tab:table-1}
%\begin{minipage}{140pt}%{<preferred-table-width>}
\begin{adjustbox}{width=\columnwidth,center}
\begin{tabular}{|c||c|c|c|}
	\hline
	local physics laws & Special Relativity & Electromagnetism & Thermodynamics \\
	\hline \hline
	quantities & $m = m_0 a^{-b/2}$ & $e = e_0 a^{-b/4}\,, \lambda = \lambda_0 a \,, \nu = \nu_0 a^{-1+b/4}$ & $T = T_0 a^{-1}$ \\
	\hline
	constants & $c = c_0 a^{b/4} \,, G = G_0 a^{b}$ & $\epsilon = \epsilon_0 a^{-b/4} \,, mu = \mu_0 a^{-b/4} $ & $k_{\textrm{B} 0} \,, \hbar = \hbar_0 a^{-b/4}$ \\
	\hline
	energies & $m c^2 = m_0 c_0^2$ & $h \nu = h_0 \nu_0 a^{-1}$ & $k_{\textrm{B}} T = k_{\textrm{B}} T_0 a^{-1}$ \\
	\hline
\end{tabular}
\end{adjustbox}
%\end{minipage}
\end{table}

\section{The 3+1 formalism of the meVSL model for the RW metric} 
\label{sec:ADM}

The 3+1 formalism in GR involves decomposing four-dimensional spacetime into three-dimensional spatial hypersurfaces and one temporal dimension, allowing for the analysis of spacetime evolution. This process begins by embedding a three-dimensional hypersurface, $\Sigma_{\T}$, into the four-dimensional manifold of spacetime. On this hypersurface, the induced metric, denoted $\gamma_{ij}$, is derived from the spacetime metric $g_{\mu\nu}$. The induced metric describes distances and angles within the hypersurface and is obtained by projecting the spacetime metric onto the hypersurface.

To facilitate this projection, a projection operator $p^{\mu}_{~\nu}$, is used to separate the tangential and orthogonal components of vectors and tensors relative to the hypersurface. The projection operator is defined as $p^{\mu}_{~\nu} = \delta^{\mu}_{~\nu} + n^{\mu} n_{\nu}$, where $n^{\mu}$ is the unit normal vector to the hypersurface. Therefore, vectors in spacetime can thus be decomposed into components tangential to the hypersurface, which are affected by the induced metric $\gamma_{ij}$, and components orthogonal to the hypersurface, aligned with the normal vector $n^{\mu}$. 

The induced covariant derivative, $D_i$, is defined on the hypersurface using the induced metric, allowing for the computation of derivatives of tensor fields confined to the hypersurface while maintaining compatibility with the metric. The curvature of the hypersurface has two characters: intrinsic curvature, $R_{ij}$, which describes the internal curvature of the hypersurface itself and derived from the Riemann curvature tensor associated with $\gamma_{ij}$, and extrinsic curvature, $K_{ij}$, which describes how the hypersurface is embedded within the higher-dimensional spacetime. Extrinsic curvature measures the change of the normal vector as it moves along the hypersurface and is crucial for understanding the hypersurface's temporal evolution, given by $K_{ij} = -\frac{1}{2} \mathcal{L}_n \gamma_{ij}$, where $\mathcal{L}_n$ is the Lie derivative along the normal vector $n^{\mu}$.

\subsection{Embedding a hypersurface into a manifold}
\label{subsec:hyper}

%Best notation follow 2210.10103v2. \\
We consider a 4-dimensional manifold $\mathcal{M}$ with a metric $\tf{g} = g_{\mu\nu} \, \of{\theta}^{\mu} \otimes \of{\theta}^{\nu} \equiv g_{\mu\nu} \, \of{\theta}^{\mu} \of{\theta}^{\nu}$, where it is common to omit the Cartesian product signs $\otimes$. In this manuscript, we define the number of straight lines above a character, $l$, as representing a rank $(l,0)$ tensor and the number of wavy lines below a character, $m$, as representing a rank $(0,m)$ tensor. $\of{\theta}^{\mu}$ is intended to represent the coordinate basis one-forms $dx^{\mu}$ as we use $\bar{e}_{\mu}$ to reprent $\partial_{\mu}$. That is, we mean $\tvec{T} = T^{\mu\nu} \ovec{e}_{\mu} \ovec{e}_{\nu}$ and $\tf{g} = g_{\mu\nu} \of{\theta}^{\mu} \of{\theta}^{\nu}$. Let $x^{\mu}$ be the coordinates on this manifold. $x^{\mu} = X^{\mu}(\xi^i)$ is the definition of the embedding of a 3-dimensional spacelike hypersurface $\Sigma$, where $\mu = 0, 1, 2, 3$ and $i = 1, 2, 3$.  There must be a one-to-one mapping $X: \Sigma \rightarrow \mathcal{M}$. When the line element is limited to the hypersurface $\Sigma$, the metric on $\mathcal{M}$ produces a metric on $\Sigma$
\begin{align}
ds^2 &= g_{\mu\nu} (x^{\lambda}) dx^{\mu} dx^{\nu} \Big|_{\Sigma} = g_{\mu\nu} (X^{\lambda}) \frac{\partial X^{\mu}}{\partial \xi^{i}} d\xi^{i}  \frac{\partial X^{\nu}}{\partial \xi^{j}} d\xi^{j} \equiv g_{\mu\nu} X^{\mu}_{i} X^{\nu}_{j} d\xi^{i} d\xi^{j} \equiv \gamma_{ij} d\xi^{i} d\xi^{j} \label{ds2onm} \,,
\end{align}
where us define $X^{\mu}_{i} \equiv \partial X^{\mu}/ \partial \xi^{i}$ to be the $\mu$-th component along $x^{\mu}$ of the $i$-th vector in the natural basis over $\Sigma$ given by $\ovec{e}_{i} \equiv \partial / \partial \xi^{i}$.  It follows 
\begin{align}
X^{\mu}_{i} \ovec{e}_{\mu} \equiv  X^{\mu}_{i} \partial_{\mu} = \frac{\partial X^{\mu}}{\partial \xi^{i}} \frac{\partial}{\partial X^{\mu}} = \frac{\partial}{\partial \xi^{i}} \equiv \partial_{i} \equiv \ovec{e}_{i} \label{ei} \,.
\end{align}
We then have a natural definition of the induced metric $\tf{\gamma}$ on $\Sigma$ given by
\begin{align}
\tf{\gamma} &\equiv \gamma_{ij} \of{\theta}^{i} \of{\theta}^{j} = g_{\mu\nu} X^{\mu}_{i} X^{\nu}_{j}  \of{\theta}^{i} \of{\theta}^{j} =  \left( \frac{\partial}{\partial \xi^{i}} \cdot  \frac{\partial}{\partial \xi^{j}} \right) \of{\theta}^{i} \of{\theta}^{j}  \label{gammaab} \,.
\end{align}
%\begin{align}
%\tf{\gamma} &\equiv \gamma_{ij} \of{\theta}^{i} \of{\theta}^{j} = g_{\mu\nu} X^{\mu}_{i} X^{\nu}_{j}  \of{\theta}^{i} \of{\theta}^{j}  = g_{\mu\nu} \frac{\partial X^{\mu}}{\partial \xi^{i}}  \frac{\partial X^{\nu}}{\partial \xi^{j}}  \of{\theta}^{i} \of{\theta}^{j} = \ovec{e}_{\mu} \cdot \ovec{e}_{\nu} \frac{\partial X^{\mu}}{\partial \xi^{i}}  \frac{\partial X^{\nu}}{\partial \xi^{j}}  \of{\theta}^{i} \of{\theta}^{j} \nonumber \\ 
%&=  \left( \frac{\partial}{\partial X^{\mu}} \cdot  \frac{\partial}{\partial X^{\nu}} \right) \frac{\partial X^{\mu}}{\partial \xi^{i}}  \frac{\partial X^{\nu}}{\partial \xi^{j}}  \of{\theta}^{i} \of{\theta}^{j} =  \left( \frac{\partial}{\partial \xi^{i}} \cdot  \frac{\partial}{\partial \xi^{j}} \right) \of{\theta}^{i} \of{\theta}^{j}  \label{gammaab} \,.
%\end{align} %This expression is consistent with Lyder Eq.(3.45), (3.47), (3.52)
The tangent space $T_p \Sigma$, representing the manifold $\Sigma$ at point $p$, is formed using the three vectors $\ovec{e}_i$ as a basis. This space is a subspace of $T_p \mathcal M$, the tangent space to ${\mathcal M}$. We build the orthogonal complement to $T_p \Sigma$, defined by the metric $\tf{g}$, to finish the foundation of this space. The vector orthogonal to $\ovec{e}_i$ represented as $\ovec{n}$ yield this subspace. This normal vector, with components $n^{\mu}$ in the $\partial_{\mu}$ basis, satisfies 
\begin{align}
g_{\mu\nu} X^{\mu}_{i} n^{\nu} = 0 \quad , \quad \tf{g} \left( \ovec{n}\,,\ovec{n} \right) = g_{\mu\nu} n^{\mu} n^{\nu} = n^{\mu} n_{\mu} = -1 \label{Xmuanmu} \,,
\end{align}
where $-1$ is because the $\Sigma$ is spacelike and we adopt the \((-\,,+\,,+\,,+)\) metric signature convention. These two conditions fully determine the vector $\ovec{n}$.  We then have the set of vectors $(\ovec{e}_i, \ovec{n})$ forming a basis of $T_p \mathcal M$ for each point $p$. With them, we can express any vector of $T_p \mathcal M$ as a linear combination of the basis in hypersurface $\Sigma$ and the vector that is normal to it in the following way
\begin{align}
\left( \ovec{A} \right)^{\mu} = \left( A^{\perp} \ovec{n} + A^i \ovec{e}_i \right)^{\mu} \equiv  A^{\perp} n^{\mu} + A^i X^{\mu}_{i} \equiv  - \left( \ovec{A} \cdot \ovec{n} \right) n^{\mu} + A^i X^{\mu}_{i} \label{Amu} \,.
\end{align}
By using Eq.~\eqref{ei}, the spatial components of the vector can be rewritten as 
\begin{align}
A^{i} = \gamma^{ij} A_{j} = \gamma^{ij} \left( \ovec{A} \cdot \ovec{e}_j \right) = \gamma^{ij} g_{\mu\nu} X^{\mu}_{j} A^{\nu} \equiv X^{i}_{\nu} A^{\nu} \label{Ai} \,.
\end{align}
%\begin{align}
%A^{i} = \gamma^{ij} A_{j} = \gamma^{ab} \left( \ovec{A} \cdot \ovec{e}_b \right) =  \gamma^{ab} g_{\mu\nu} A^{\nu} \left( \ovec{e}_{b} \right)^{\mu} \equiv \boxed{ \gamma^{ab} g_{\mu\nu} X^{\mu}_{b} } A^{\nu} = X^{a}_{\nu} A^{\nu} \label{Aa} \,.
%\end{align}
This operator, $X^{i}_{\nu}$, allows us to find the $i$-th component of vector $\ovec{A}$.  From Eq.~\eqref{Amu}, we obtain
\begin{align}
X^{\mu}_{i} A^{i} = \left( \ovec{A} \right)^{\mu} + \left(  \ovec{A} \cdot \ovec{n} \right) n^{\mu} = A^{\nu} \delta^{\mu}_{\nu} + A^{\nu} n_{\nu} n^{\mu} = \left(  \delta^{\mu}_{\nu} + n^{\mu} n_{\nu} \right) A^{\nu} \equiv p^{\mu}_{\nu} A^{\nu} \label{XmuaAa} \,.
\end{align}
Therefore, we have found the projection operator, $p_{\mu\nu}$, that commands a vector of spacetime to one on the hypersurface. We emphasize that $X^{\mu}_i$ acts on $3$-vectors and $p^{\mu}_{~\nu}$ ($X_{\mu}^{i}$) acts on $4$-vectors. The properties of the projection operator are
\begin{align} 
p^{\mu}_{~\nu} \, n^{\nu} = 0 \qquad , \qquad p^{\mu}_{~\rho} \, p^{\rho}_{~\nu} = p^{\mu}_{~\nu} \qquad , \qquad p^{\mu}_{~\mu} = 3 \,. \label{pproperties} 
\end{align}
The orthogonal projector onto $\Sigma$ is the operator $\vec{p}$ associated with this orthogonal decomposition of $T_p (\mathcal M)$ into $T_p (\Sigma)$ and the one-dimensional subspace of the manifold generated by the normal vector $\ovec{n}$. We can also define an inverse mapping $\vec{p}^{~\ast}_{\M} : T^{\ast}_p (\Sigma) \rightarrow T^{\ast}_p (\M)$.  We adopt this operator when we link three-dimensional covariant derivatives with four-dimensional ones.

\subsection{Induced covariant derivaties}
\label{subsec:ICD}

The covariant derivative defined by the metric $\tf{g}$ of a vector in $\Sigma$, in the direction of another vector also in $\Sigma$, is generally not a tangent vector to the hypersurface. To define a covariant derivative on $\Sigma$ naturally, it must remain a vector in $T_p \Sigma$. Consider the case where the first three coordinates $x^{\mu}$ match $\xi^{i}$, (\textit{i.e.},  $x^{i} = \xi^{i}$). These adapted coordinates (ACs) simplify the analysis since the vectors $\ovec{e}_i$ align with the coordinate basis of $T_p \mathcal{M}$, making their components trivial: $X_{i}^{\mu} = \delta_{i}^{\mu}$. Using these ACs, we can express the covariant derivative of a vector $\ovec{A} = A^i \ovec{e}_i$ on the hypersurface in the direction of the basis vectors as 
\begin{align}
             ^4\nabla_{\ovec{e}_i} \ovec{A} &\equiv ^4\nabla_{i} \ovec{A} = \left(  \frac{\partial A^k}{\partial X^i} + \Gamma^{k}_{~ij} A^j \right) \ovec{e}_{k} + ^4\Gamma^{0}_{~ij} A^j \ovec{e}_{0} \nonumber \\
	  &\equiv D_{i} A^{k} \ovec{e}_{k} + ^4\Gamma^{0}_{~ij} A^j \ovec{e}_{0} \equiv A^{k}_{|i} \ovec{e}_{k} + ^4\Gamma^{0}_{~ij} A^j \ovec{e}_{0} \label{DaAc} \,.
\end{align}
%\begin{align}
%             ^4\nabla_{\ovec{e}_a} \ovec{A} &\equiv ^4\nabla_{a} \ovec{A} =^4 \nabla_{a} \left( \ovec{e}_b A^b \right)
%             =\ovec{e}_b \frac{\partial A^b}{\partial X^a}+ \left(^4\Gamma^{\mu}_{~ab}\ovec{e}_{\mu} \right)A^b \nonumber \\
%             &=\ovec{e}_b \frac{\partial A^b}{\partial X^a}+ \left(^4\Gamma^{c}_{~ab}\ovec{e}_{c}+^4\Gamma^{0}_{~ab} \ovec{e}_{0} \right)A^b = \left(  \frac{\partial A^c}{\partial X^a} + \Gamma^{c}_{~ab} A^b \right) \ovec{e}_{c} + ^4\Gamma^{0}_{~ab} A^b \ovec{e}_{0} \nonumber \\
%	  &\equiv D_{a} A^{c} + ^4\Gamma^{0}_{~ab} A^b \ovec{e}_{0} \equiv A^{c}_{|a} + ^4\Gamma^{0}_{~ab} A^b \ovec{e}_{0} \label{DaAc} 
%\end{align}
If we now consider the $k$-th component of the covariant derivative of vector $\ovec{A}$ in the direction of $\ovec{e}_i$, it will be of the form:
\begin{equation}
    \left( D_i \ovec{A} \right)_k \equiv A_{k|i} =\ovec{e}_k \cdot D_{\ovec{e}_i} \ovec{A} \equiv A_{k,i}-\Gamma_{jki}A^j \label{DaAc2} \,,
\end{equation}
where 
\begin{equation}
    \Gamma_{ijk} \equiv ^3\Gamma_{ijk}= \ovec{e}_i \cdot \nabla_k \ovec{e}_j = \frac{1}{2}\left(\gamma_{ij,k}+\gamma_{ik,j}-\gamma_{jk,i}\right) \label{3Gammaabc} \,.
\end{equation}
To facilitate the calculations, let us note that the covariant derivative can also be expressed (for $A^{\mu}=A^jX_{j}^{\mu}$,  $A^{j}=A^{\mu} X^{j}_{\mu}$) in terms of components as follows,
\begin{align}
    A_{~|k}^{j} = (A_{~;\nu}^{\mu}X^{\nu}_{k})^j =X_{\mu}^{j}(A_{~;\nu}^{\mu}X^{\nu}_{k}) =X_{\mu}^{j}X_{k}^{\nu}A^{\mu}_{~;\nu} \label{Akj} \,.
\end{align}
For a given tensor field $\mathbf{T}$ on the hypersurface $\Sigma$, its covariant derivative for the induced metric $\tf{\gamma}$ can be expressed in terms of the covariant derivative $^{4}\nabla \mathbf{T}$ for the spacetime metric $\tf{g}$. This relationship can be expressed as
\begin{align}
\of{D} \mathbf{T} = \vec{p}^{~\ast} \,^{4} \of{\nabla} \mathbf{T} \label{DT} \,.
\end{align}
In terms of components, this relationship is given by 
\begin{align}
D_\rho T^{\alpha_1\ldots\alpha_p}_{\ \qquad\beta_1\ldots\beta_q}
= p_{\ \ \, \mu_1}^{\alpha_1} \, \cdots
p_{\ \ \, \mu_p}^{\alpha_p} \,
p_{\ \ \, \beta_1}^{\nu_1} \, \cdots
p_{\ \ \, \beta_q}^{\nu_q} \,
p_{\ \ \, \rho}^{\sigma} \, \nabla_\sigma
T^{\mu_1\ldots\mu_p}_{\ \qquad\nu_1\ldots\nu_q}  .  \label{link_D_nab_comp}
\end{align}
It shows how the intrinsic covariant derivative of a tensor on the hypersurface relates to its spacetime counterpart.

%In the later parts of this manuscript, we consider only the \textbf{case of adapted coordinates}. \textcolor{blue}{Important summary \begin{align} X^{\mu}_{i} A^{i} &= \left( \delta^{\mu}_{\nu} + n^{\mu} n_{\nu} \right) A^{\nu} \equiv p^{\mu}_{\nu} A^{\nu} \label{Amusum} \\ X^{i}_{\mu} A^{\mu} &= A^{i}  \label{Aasum} \\ A^{j}_{|k} &= \left( D_{k} A \right)^{j} \overset{\textrm{Eq}.~\eqref{Aasum}}{=} X^{j}_{\mu} \left( D_{k} A \right)^{\mu} \overset{\textrm{Eq}.~\eqref{Amusum}}{=} X^{j}_{\mu} X^{\nu}_{k} \nabla_{\nu} A^{\mu} \label{DcAb} \end{align} }

\subsection{Lapse and shift functions}
\label{subsec:ADM}

%Decomposition to $3+1$ Sec.3.4, Arnowitt–Deser–Misner (ADM) \\ 
%In this section, we describe formulations of Einstein’s equation based on the spacetime foliation, i.e., ADM and $N+1$ formulations.
We have demonstrated how to embed a three-dimensional geometry into spacetime by introducing the metric $\gamma_{ij}$ and defining an appropriate covariant derivative on the hypersurface that allows for parallel transport. This framework enables us to decompose spacetime vectors into components orthogonal and tangential to the hypersurface. A complete covering of spacetime by non-intersecting hypersurfaces is called foliation. In this context, a function $x^0 \equiv ct \equiv \T$ in spacetime describes the foliation, where each hypersurface corresponds to a constant value of $\T$ (\textit{i.e.}, $\T = \text{const}$). In the traditional approach, since the speed of light $c$ is considered constant, $x^0 = ct = \text{constant}$, can be identified with the case where $t$ is constant. However, if we include the possibility that the speed of light $c$ changes over cosmic time,  $x^0$ should be regarded as constant, meaning that in cases where the speed of light varies cosmologically,  $c[t_i] t_i$ remains constant, allowing $x^0$ to foliate constant hypersurfaces. On a given hypersurface, labeled by $\mathcal{T} = \mathcal{T}_0$, an event at a point $p$ is characterized by three spatial coordinates $\xi^i$. This point $p$ is mapped to a point $p'$ on the next hypersurface $\mathcal{T}_0 + d\mathcal{T}$. While $p'$ shares the same intrinsic coordinates $\xi^i$ as $p$, its spacetime coordinates differ as a function of $\mathcal{T}$. The functions in the following equation provide an analytical description of this foliation
\begin{align}
    x^{\mu}=X^{\mu}(\T\,,\xi^{i}) \overset{\textrm{ACs}}{\rightarrow} X^{\mu}(x^0\,,x^{i}) = X^{\mu}(ct\,,x^{i})  \label{xmu2} \,,
\end{align}
where we call it as an adapted coordinates (ACs) when $\xi^i = x^i$. And from now on, we will continue to use these ACs throughout this manuscript.

Let $\ovec{\T}$ be a timelike vector field on the spacetime tangent to the time axis, represented as $\ovec{\T} = \partial / \partial \T$. This vector field satisfies the condition $\nabla_{\T} \T = \T^{\alpha} \nabla_{\alpha} \T = 1$. Note that $\ovec{\T}$ is not always normal to the hypersurface $\Sigma_{\T}$; it has components both tangent to $\Sigma_{\T}$ and along the normal vector $\ovec{n}$. The vector $\ovec{\T}$ represents the connection between points on two different hypersurfaces. It can be decomposed into its projection onto the hypersurface $\Sigma_{\T}$ and its component along the normal vector $\ovec{n}$
\begin{align}
\bar{{\T}} &\equiv \frac{\partial}{\partial \T} \equiv \T^{\perp} \bar{n} + \T^{i} \bar{e}_i \equiv \T^{\alpha} \bar{e}_{\alpha} \label{barT} \,,
\end{align}
where
\begin{align}
\T^{i} &= \gamma^{ij} \T_{j} = \gamma^{ij} g_{\mu\nu} X_{j}^{\nu} \T^{\mu} \equiv X^{i}_{\mu} \T^{\mu} \label{Ti}  \quad \textrm{and} \\
\T^{\alpha} &= -\left( \bar{\T} \cdot \bar{n} \right)  n^{\alpha} + X_{\mu}^{i} \T^{\mu} X^{\alpha}_{i} \equiv N n^{\alpha} + N^i X^{\alpha}_i \label{Talpha} \,.
\end{align}
The scalar $N$ is the lapse function, and the shift vector is the vector $N^i$ on the hypersurface. Along with the metric $\gamma_{ij}$, these constitute the ADM variables. The lapse function $N$ determines the proper time interval between neighboring spatial hypersurfaces, effectively controlling the rate at which time progresses at each point in space. It sets the arrangement of spatial hypersurfaces within the spacetime manifold, a configuration known as the time-slicing condition. The shift vector $N^i$, on the other hand, determines the spatial direction of the time axis at each point, influencing how the spatial coordinates shift on hypersurfaces as time evolves. This choice affects the positioning of spatial points on the hypersurfaces and is known as the spatial gauge condition. Specifying the four functions $N^{\mu}$ fully determines the foliation of hypersurfaces that generate the spacetime structure. To finalize this determination, we need to find the metric induced on each hypersurface and derive it from the dynamical equations.  

 In this manuscript, we aim to apply the $3+1$ formalism to the RW metric with a VSL in a cosmological context. To achieve this, we will first briefly review the definition of Eulerian observers and the characteristics of the RW metric, then derive the corresponding ADM variables.

\subsection{Eulerian observers}
\label{subsec:Eulerian}

The worldlines of Eulerian observers are orthogonal to the hypersurfaces in the context of the $3+1$ formalism. In the $3+1$ formalism, spacetime is split into a series of spatial hypersurfaces $\Sigma_{\T}$ that represent slices of space at a constant $\T = ct$. An Eulerian observer remains at a fixed position in space on these hypersurfaces (\textit{i.e.}, at rest relative to $\Sigma_{\T}$). The worldline of such an observer is the path they trace out in spacetime as time progresses, which is purely along the time direction. Their proper time is generally $N dt$, which might differ from the coordinate time depending on the value of the lapse function $N$.

The point is that these worldlines are orthogonal to the spatial hypersurfaces $\Sigma_{\T}$. This orthogonality arises because the Eulerian observer's four-velocity is aligned with the normal vector to the hypersurfaces. The worldline of an Eulerian observer is along $\ovec{n}$, meaning the observer is moving through time but not through space (remaining at a fixed spatial position).

The unit vector $\ovec{n}$ normal to the $\Sigma$ is defined only at points belonging to $\Sigma_{\T}$. Thus, the worldlines of Eulerian observers are orthogonal to the hypersurfaces $\Sigma_{\T}$s. It means that each hypersurface $\Sigma_{\T}$ represents a set of events that are simultaneous from the perspective of an Eulerian observer.

The 4-acceleration $a$ of an Eulerian observer is defined as:
\begin{align}
\ovec{a} = \nabla_{\ovec{n}} \ovec{n} \label{oveca} \,.
\end{align}
It has been noted that $\ovec{a}$ is orthogonal to $\ovec{n}$ and, therefore, tangent to $\Sigma_{\T}$. Because the Eulerian observers are hypersurface-orthogonal, the congruence formed by their worldlines has zero vorticity, which is why they are sometimes called non-rotating observers. 

\subsection{The RW metric of the meVSL model}
\label{subsec:RW}

In cosmological models, a foundational assumption is Weyl's postulate. This postulate asserts that a bundle of geodesics in spacetime possesses space-like hypersurfaces orthogonal to these geodesics. With the help of this framework, it is possible to include a temporal parameter $t$, which acts as a gauge for the proper time along these geodesics and guarantees consistency in the description of temporal evolution throughout the model. The proper time is the time interval measured by a clock moving along a worldline in spacetime. To effectively capture the spatial aspects of this model, we introduce spatial coordinates $(x^1, x^2, x^3)$, which remain constant along any given geodesic. Each galaxy can be assigned fixed coordinates in this spatial framework (\textit{i.e.}, comoving with the cosmic expansion), reflecting the assumption that galaxies do not move spatially relative to the expanding universe \cite{Islam01,Narlikar02,Hobson06,Roos15} %Use Islam p.38-39
\begin{align}
x^{\mu}(\tau) &= \left( x^0 = c \tau\,,x^1=\textrm{const}\,,x^2=\textrm{const}\,,x^3=\textrm{const} \right) \label{xmutau} \\
x^{\mu}(t) &= \left( x^0 = c t\,,x^1=\textrm{const}\,,x^2=\textrm{const}\,,x^3=\textrm{const} \right) \label{xmut} \,.
\end{align}

\begin{figure*}
\centering
\vspace{1cm}
%\begin{tabular}{cc}
\includegraphics[width=1\linewidth]{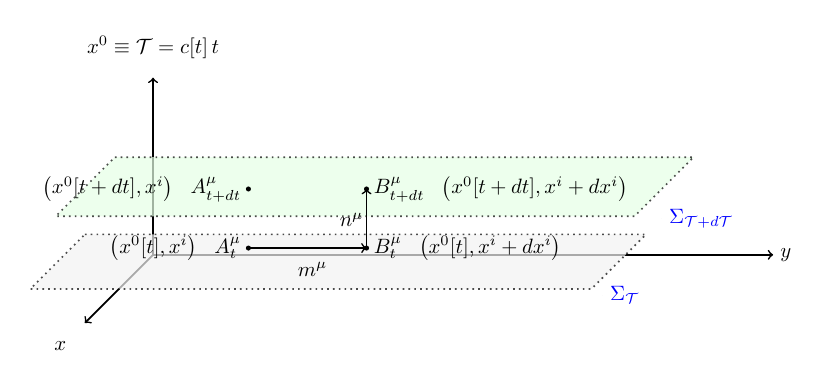} 
\vspace{-0.5cm}
\caption{A foliation of the spacetime of the RW metric. The hypersurfaces $\Sigma$s are level surfaces of the temporal coordinate $\T$s. The normal vector $\ovec{n} = n^{\mu} \ovec{e}_{\mu}$ is orthogonal to these $\T = \textrm{const}$ spatial hypersurfaces. } \label{fig-2}
\vspace{1cm}
\end{figure*}

In the SMC, we consider the speed of light to be constant. Therefore, we treat the differential $dx^0$ as infinitesimally small changes in time multiplied by $c$ (\textit{i.e.} $dx^0 = c dt$). However, in the meVSL model, where $c$ is a function of cosmic time, $dx^0$ is regarded as the approximate change of the function $x^0$ and is described as \cite{Lee:2020zts,Lee:2022heb,Lee:2023bjz,Lee:2024part}
\begin{align}
dx^0 \equiv d \T = d \left(c[t] t \right) = \left( 1 + \frac{d \ln c[t]}{d \ln a} \frac{d \ln a}{d \ln t} \right) c[t] dt = \left( 1 + \frac{d \ln c[t]}{d \ln a} H t \right) c[t] dt \equiv \tilde{c}[t] dt \label{dx0} \,,
\end{align} 
where $H$ is the Hubble parameter.
The infinitesimal spacetime interval under these assumptions can be expressed as
\begin{align}
ds^2 = -\tc^2 dt^2 + \gamma_{ij} dx^i dx^j = -\tc^2 dt^2 + a^2(t)\sigma_{ij}(\vec{x}) dx^i dx^j  \label{ds2} \,,
\end{align}
where $\gamma_{ij}$ are functions of $(ct, x^1, x^2, x^3)$, reflecting the dynamic nature of space as influenced by the evolving universe.  %\textcolor{blue}{Differentials, such as $dt$, are specific cases of one-forms and represent infinitesimally small changes in a variable. They are often written in the form $df$,  where $f$ is a function, and $df$ represents the differential of $f$. For a function $f(x)$, the differential $df$ is given by $df = (\partial f/\partial x)dx$, which describes how $f$ changes as $x$ changes. Differentials are used to describe changes in variables and to compute derivatives. In calculus, $df$ is used to approximate changes in a function. They play a key role in defining the gradient of a function and in the theory of integration. Galaxies in the RW metric are comoving with the cosmic expansion. They are not accelerating or moving through the spatial hypersurfaces; they just follow the expansion of the universe. Their proper time is the cosmological time $t$ in the RW metric, which is the same as the coordinate time.  While Eulerian observers in the 3+1 formalism are at rest relative to the spatial slices, their proper time is generally $Ndt$, which might differ from the coordinate time depending on the value of the lapse function.}
The construction of this metric ensures that the proper time $\tau$ along a galaxy's worldline is identical to the coordinate time $t$, as the spatial components vanish $dx^i = 0$ along these geodesics \cite{Islam01,Narlikar02,Hobson06,Roos15}. This is shown in Fig.~\ref{fig-2}. Consequently, the metric simplifies to $ds = \tc dt = \tc d \tau$, directly correlating the proper time with the coordinate time. This setup underscores the orthogonality of the galaxy worldlines to the hypersurfaces of constant time. Specifically, a vector $A^{\mu} = (\tc\, dt, 0, 0, 0)$ along the worldline is orthogonal to a vector $B^{\mu} = (0, dx^1, dx^2, dx^3)$ within the hypersurface at constant $\T$ (\textit{i.e.}, $g_{\mu\nu} A^{\mu}B^{\nu} = 0$). Such a model provides a clear and coherent structure for analyzing cosmological phenomena and serves as a robust foundation for further investigations into the dynamics of the universe. As we saw in the previous subsection~\ref{subsec:Eulerian}, each galaxy in the RW metric can be identified as an Eulerian observer. The worldline given by Eq.~\eqref{xmutau} satisfies the geodesic equation
\begin{align}
\frac{d^2x^{\mu}}{ds^2} + \Gamma^{\mu}_{\nu\lambda} \frac{d x^{\nu}}{ds} \frac{dx^{\lambda}}{ds} = 0 \label{geodesicEq} \,,
\end{align} 
where we use 
\begin{align}
\frac{d x^{\mu}}{ds} = \frac{d}{\tc d\tau} \left(c \tau\,,x^i \right) = \left(\frac{\tc d \tau}{\tc d\tau}\,,\frac{d x^i}{\tc d \tau} \right) = \left(1\,,0\,,0\,,0 \right) \label{dxmuds} \,.
\end{align}

%%%%%%%%%%%%%%%%%%%%%%%%%%%%%%%%%%%%%%%
%%%%%%%%%%%%%%%%%%%%%%%%%%%%%%%%%%%%%%%

\subsection{The lapse function and induced metric}
\label{subsec:ADMRWmeVSL}

First, the RW metric is given by 
\begin{align}
g_{\mu\nu} = \text{diag} \left(-1 , \frac{a^2}{1-kr^2} , a^2 r^2, a^2 r^2 \sin^2 \theta \right)  %\begin{pmatrix}
					%-1 & 0 & 0 & 0 \\
				%	0 & \frac{a^2}{1-kr^2} & 0 & 0 \\
					%0 & 0  & a^2 r^2 & 0 \\
					%0 & 0 & 0 & a^2 r^2 \sin^2 \theta
				 % \end{pmatrix}  \quad 
				 \,. \label{FLRWmetric} 	
\end{align}
In this subsection, we aim to express the RW metric in ADM variables within the meVSL model, where the speed of light varies. To do so, we decompose the infinitesimal line element between two spacetime events, $p(x^0\,,x^i)$ and $p'(x^0+dx^0\,,x^i + dx^i)$ into two parts as \cite{MTW:17}
\begin{align}
ds^2 &= - \begin{pmatrix} \textrm{change in $x^0$ from lower} \\ \textrm{to upper hypersurface} \end{pmatrix}^2 + \begin{pmatrix} \textrm{proper distance} \\ \textrm{in lower hypersurface} \end{pmatrix}^2 \nonumber \\ 
	&= - \left( N[t] c[t] dt \right)^2 + g_{ij} \left( dx^{i} + N^{i}\left( t,\vec{x} \right) c[t] dt \right) \left( dx^{j} + N^{j}\left( t,\vec{x} \right) c[t] dt  \right) \nonumber \\ 
	%&= - N^2 \left( 1 -\frac{N_k N^{k}}{N^2} \right) c^2 dt^2 + 2 N_i dx^i c dt + g_{ij} dx^i dx^j  \nonumber \\
	%&= - N^2 \frac{c^2}{\tc^2} \left( 1 -\frac{N_k N^{k}}{N^2} \right) \tc^2 dt^2 + 2 N_i \frac{c}{\tc} dx^i \tc dt + g_{ij} dx^i dx^j  \nonumber \\
	&= - \left( 1 -\frac{N_k N^{k}}{N^2} \right) \tc^2 dt^2 + 2 \frac{N_i}{N}  dx^i \tc dt + g_{ij} dx^i dx^j  \label{ds22} \\
	&\overset{\textrm{RW}}{=} - \tilde{c}^2 dt^2 + a^{2}(t) \sigma_{ij} dx^i dx^j \equiv g_{\mu\nu} dx^{\mu} dx^{\nu} \qquad \textrm{for the RW metric} \label{ds23}  \,,
\end{align}
where we define the lapse function $N \equiv \tc/c$ as
\begin{align}
x^{0}(t+dt) - x^{0}(t) &=  \left( c[t+dt] \right) \left( t+dt \right) - c(t) t \overset{\mathcal{O}(1)}{\approx} c[t] dt + dc[t] t \nonumber \\
	&= \left( 1 + \frac{d \ln c}{d \ln t} \right) c[t] dt \equiv \tilde{c}[t] dt \equiv N[t] c[t] dt \label{NmeVSL} \,,
\end{align}
where we also use Eq.~\eqref{dx0}. Thus, we can obtain the components of the metric and the inverse metric from Eq.~\eqref{ds22} as
\begin{align}
g_{\mu\nu} &= \left(
\begin{array}{c|c} g_{00} & g_{0j} \\ \hline
g_{i0} & g_{ij} \\
\end{array}
\right) = \left(
\begin{array}{c|c} (- N^2 + N_k N^{k} ) c^2/\tc^2 & N_j c/\tc  \\ \hline
N_i c/\tc& \gamma_{ij} \\
\end{array}
\right) = \left(
\begin{array}{c|c} (- 1 + N_k N^{k} /N^2) & N_j /N  \\ \hline
N_i /N & \gamma_{ij} \\
\end{array}
\right) \nonumber \\ 
&\overset{\textrm{RW}}{=} \left(
\begin{array}{c|c} 
- 1 & 0 \\ \hline
0 & a^{2}(t) \sigma_{ij} \\
\end{array}
\right) \equiv - n_{\mu} n_{\nu} + \gamma_{\mu\nu} = \bar{{e}}_{\mu} \cdot \bar{{e}}_{\nu} \label{gmunnuRW} \,, \\  %Shapiro p.2 Eq1.3 
g^{\mu\nu} &= 
 \frac{\tc^2}{N^2 c^2} \left(
\begin{array}{c|c} 
-1 & N^{j} c/\tc\\ \hline
N^{i} c/\tc & \left( N^2 g^{ij} - N^{i} N^{j} \right) c^2/\tc^2 \\
\end{array}
\right) = 
\left(
\begin{array}{c|c} 
-1 & N^{j} / N \\ \hline
N^{i} / N & \left( g^{ij} - N^{i} N^{j}/N^2 \right) \\
\end{array}
\right)  \nonumber \\ 
&\overset{\textrm{RW}}{=} \left(
\begin{array}{c|c} 
- 1 & 0 \\ \hline
0 & a^{2}(t) \sigma^{ij} \\
\end{array}
\right) \equiv \of{\theta}^{\mu} \cdot \of{\theta}^{\nu} 
\label{gmunform} \,. \end{align}  %Shapiro Eq1.8 correct one
In this work, \( \of{\theta}^\mu \) represents a basis of one-forms, and the operation \( \cdot \) is explicitly defined as the bilinear form induced by the metric \( g_{\mu\nu} \). Specifically, definition~\eqref{gmunform} is consistent with the metric structure and allows us to contract one-forms in a way that recovers the components of the inverse metric.

The $1$-form dual to the normal timelike vector $\ovec{n}$ defined only at points belonging to is given by
\begin{align}
&\of{n} = n_{\beta} \of{\theta}^{\beta} = n_{0} c[t] dt + 0 + 0 + 0 \quad , \quad n_{0} c dt \equiv - c dt  \nonumber\\
&n_{0} = - 1 \quad \Rightarrow \quad n_{\beta} = \left( -1\,,0\,,0\,,0 \right) \label{ofn} \,.
\end{align}
Then, we can obtain components of the normal vectors as
\begin{align}
n^{\alpha} = g^{\alpha\beta} n_{\beta} = \begin{pmatrix} 1 \\ \vec{0} \end{pmatrix} \label{nalpha2} \,.
\end{align} 
The normal vector is
\begin{align}
&\ovec{n} = n^{\alpha} \ovec{e}_{\alpha} = n^{0} \frac{\partial}{c \partial t} + n^{i} \frac{\partial}{\partial x^i} \quad , \quad
n^{\alpha} = \left(n^0\,, n^i \right) = \left( 1\,, -\frac{N^i}{N} \right) \label{noalpha} \\ 
&g_{\alpha\beta} n^{\alpha} = \left(
\begin{array}{c|c} (- 1 + N_k N^{k} /N^2) & N_j /N  \\ \hline
N_i /N & \gamma_{ij} \\
\end{array}
\right) \times \left(
\begin{array}{c} 1 \\ \hline - N^j/N \\
\end{array}
\right) = \left( -1\,, \vec{0} \right) = n_{\beta} \nonumber \\
&\langle \of{n}\,, \ovec{n} \rangle = g_{\mu\nu} n^{\alpha} n^{\beta} = -1  \label{ipnn} \,.
\end{align}
Four velocity of the Eulerian observers is given by 
\begin{align}
U^{\alpha} \equiv \frac{d x^{\alpha}}{d \tau} = \left( \tc\,, \vec{0} \right) = \tc n^{\alpha} \label{Ualphacor} \,.
\end{align} 
Thus, for the RW metric under the meVSL model, $N = \tilde{c}/c$ and $N^{i} = 0$. In the SMC, $c$ is constant,  so $N = 1$ (\textit{i.e.}, $c = \tc$). However, the VSL models allow non-constant $c$ and $N$ can vary.  In the meVSL model, $c = c_0 a^{b/4}$ and thus
\begin{align}
N[t] = 1 + \frac{b}{4} H t \label{NtmeVSL} \,.
\end{align}

\subsection{Intrinsic curvature}
\label{subsec:intrinsic}

The Riemann curvature tensor $^{(4)}\mathbf{R}$ measures the curvature of a manifold by describing how vectors are transported around closed loops. The Ricci identity provides a way to express this curvature regarding covariant derivatives. According to the Ricci identity, the $^{(4)}\mathbf{R}$ is the commutator of two covariant derivatives acting on any four-vector field and written for components as
\begin{align} 
\left(\nabla_\alpha\nabla_\beta
- \nabla_\beta\nabla_\alpha\right) A^{\mu}
= {}^{(4)}\!R^\mu_{\ \, \nu \alpha\beta} \, A^\nu \quad \textrm{for} \quad \ovec{A} \in T(\M) \label{Ricci_ident} \,.
\end{align}
This identity shows that the manifold's curvature is directly related to the non-commutativity of covariant derivatives. Thus, the Riemann tensor measures how the manifold's curvature affects the parallel transport of vectors on the manifold. The Riemann tensor is expressed in components as 
\begin{align}
{}^{(4)}\!R^\mu_{\ \, \nu \alpha\beta} = \partial_\alpha \Gamma^\mu_{\nu\beta} - \partial_\beta \Gamma^\mu_{\nu\alpha} + \Gamma^\lambda_{\nu\beta} \Gamma^\mu_{\lambda\alpha} - \Gamma^\lambda_{\nu\alpha} \Gamma^\mu_{\lambda\beta} \label{mnab} \,,
\end{align}
where $\Gamma^\mu_{\nu\beta}$ are the Christoffel symbols of the metric $\tf{g}$. The Riemann curvature tensors, Ricci tensors, and Ricci scalar curvature of the RW metric under the meVSL model are given by  
\begin{align}
&^{(4)} \tensor{R}{^0_i_0_j} = \frac{g_{ij}}{\tc^2} \left( \frac{\ddot{a}}{a} - H^2 \frac{d \ln \tc}{d \ln a} \right) , \quad
^{(4)} \tensor{R}{^i_0_0_j} = \frac{\delta^{i}_{j}}{\tc^2} \left( \frac{\ddot{a}}{a} - H^2 \frac{d \ln \tc}{d \ln a} \right) , \quad \nonumber \\
&^{(4)} \tensor{R}{^i_j_k_m} =  \left( \frac{H^2}{\tc^2} + \frac{k}{a^2} \right) \left( \delta^{i}_{k} g_{jm} - \delta^{i}_{m} g_{jk} \right) \equiv \frac{H^2}{\tc^2} \left( \delta^{i}_{k} g_{jm} - \delta^{i}_{m} g_{jk} \right) +  \tensor{R}{^i_j_k_m} \label{tRijkmApp2} \,, \\
%& \tRR_{00} = - \frac{3}{\tc^2} \left( \frac{\ddot{a}}{a} - \frac{\dot{\tc}}{\tc} H \right)  \quad , \quad
%\tRR_{ii} = \frac{g_{ii}}{\tc^2} \left( 2 \frac{\dot{a}^2}{a^2} + \frac{\ddot{a}}{a} + 2 k \frac{\tc^2}{a^2} - \frac{\dot{\tc}}{\tc} H \right) \label{tR00mp} \,, \\
&^{(4)} R_{00} = - \frac{3}{\tc^2} \left( \frac{\ddot{a}}{a} - H^2 \frac{d \ln \tc}{d \ln a} \right)  \quad ,  \nonumber \\
&^{(4)} R_{ij} = \frac{g_{ij}}{\tc^2} \left( 2 \frac{\dot{a}^2}{a^2} + \frac{\ddot{a}}{a} + 2 k \frac{\tc^2}{a^2} - H^2 \frac{d \ln \tc}{d \ln a} \right) \equiv  \frac{g_{ij}}{\tc^2} \left( 2 \frac{\dot{a}^2}{a^2} + \frac{\ddot{a}}{a} - H^2 \frac{d \ln \tc}{d \ln a} \right) + R_{ij} \label{tR00mp} \,, \\
&^{(4)} R = \frac{6}{\tc^2} \left( \frac{\ddot{a}}{a} + \frac{\dot{a}^2}{a^2} + k \frac{\tc^2}{a^2} - H^2 \frac{d \ln \tc}{d \ln a} \right) \equiv = \frac{6}{\tc^2} \left( \frac{\ddot{a}}{a} + \frac{\dot{a}^2}{a^2} - H^2 \frac{d \ln \tc}{d \ln a} \right) + R \label{tRmp2} \,.
%& \tRR = \frac{6}{\tc^2} \left( \frac{\ddot{a}}{a} + \frac{\dot{a}^2}{a^2} + k \frac{\tc^2}{a^2} - \frac{\dot{\tc}}{\tc} H \right) \label{tRmp} \,.
\end{align} 

In the context of the $3+1$ formalism, intrinsic curvature refers to the curvature inherent to a three-dimensional spatial hypersurface $\Sigma_{\T}$, as it exists independently of how it is embedded in the larger four-dimensional spacetime. A fundamental aspect of understanding spacetime slices is this curvature described by the geometry of the hypersurface. The intrinsic curvature is mathematically characterized by the Riemann curvature tensor $\mathbf{R}$ associated with the induced metric $\tf{\gamma}$. This tensor encapsulates how much the hypersurface is curved, independent of the larger spacetime it is embedded in. It implies a unique covariant derivative $D$ on the hypersurface $\Sigma_{\T}$ satisfying $D \gamma = 0$. We can define the Riemann tensor on $\Sigma_{\T}$ as  
\begin{align}
\left(D_i D_j - D_j D_i \right) A^{k} = R^k_{\ \, l i j} \, A^l \quad \textrm{for} \quad \ovec{A} \in T(\Sigma) \label{3DRicci_ident} \,.
\end{align}
The component of the Riemann tensor is expressed as 
\begin{align}
R^i_{jkl} &= \partial_k \Gamma^i_{jl} - \partial_l \Gamma^i_{jk} + \Gamma^m_{jl} \Gamma^i_{mk} - \Gamma^m_{jk} \Gamma^i_{ml} \label{Rijkl} \\
	&= \delta^{i}_{k} R_{jl} - \delta^{i}_{l} R_{jk} + \gamma_{jl} R^{i}_{k} - \gamma_{jk} R^{i}_{l} + \frac{1}{2} \left( \delta^i_l \gamma_{jk} - \delta^i_k \gamma_{jl} \right) \nonumber %Eq.(2.35) Eric
\end{align}
where $\Gamma^i_{jk}$ are the Christoffel symbols of the induced metric $\tf{\gamma}$ and $R_{jl} = R^{i}_{jil}$ is the component of Ricci tensor. The Ricci scalar $R$, a further contraction of the Ricci tensor, provides a single scalar quantity that represents the overall intrinsic curvature of the hypersurface $R = \gamma^{ij} R_{ij}$. These quantities—Riemann tensor, Ricci tensor, and Ricci scalar—provide comprehensive measures of the intrinsic curvature of the hypersurface, indicating how the hypersurface bends or curves in its own space. Then, the Riemann curvature tensors, Ricci tensors, and Ricci scalar curvature of the three-dimensional spatial part of the RW metric under the meVSL model are given by  
\begin{align}
& \tensor{R}{^i_j_k_m} =  \frac{k}{a^2} \left( \delta^{i}_{k} g_{jm} - \delta^{i}_{m} g_{jk} \right)  \label{3DtRijkmApp} \,, \\
&R_{ij} = 2 \frac{k}{a^2} g_{ij} = 2 k \sigma_{ij} \label{3DtR00mp} \,, \\
&R = 6 \frac{k}{a^2} \label{3DtRmp} \,.
\end{align}
All $3$-dimensional intrinsic curvatures do not involve the speed of light. As expected, they are identical to those of the SMC. It means that the effects of the varying speed of light are not observed on the $3$-dimensional hypersurface. From this, we can see that Lorentz invariance holds on the hypersurface. Additionally, the intrinsic curvatures are the same on all hypersurfaces, regardless of the variation in the speed of light.

\subsection{Extrinsic curvature}
\label{subsec:extrinsic}

%Extrinsic curvature: 3.3 of 2210.10103. \\
In addition to intrinsic curvature, hypersurfaces can also exhibit extrinsic curvature, which relates to their bending within the larger manifold $\mathcal{M}$. This bending describes how the normal vector $\ovec{n}$ changes direction as it moves across the hypersurface $\Sigma_{\T}$. The extrinsic curvature $K_{ij}$ of the hypersurface is defined as the projection of the covariant derivative of the normal vector $\ovec{n}$ in the direction $\ovec{e}_i$, onto the direction $\ovec{e}_j$, with a negative sign
\begin{align} K_{ij}\equiv -\ovec{e}_j \cdot ^4\nabla_i \ovec{n} \label{Kij} \,. \end{align}
Here, $K_{ij}$ is well-defined on the hypersurface because the vector $^4\nabla_i \ovec{n}$ lies in the tangent space $T_p \Sigma$. This is evident from the relation
\begin{align} ^4\nabla_i \left( \ovec{n} \cdot \ovec{n} \right)=0 \label{Dnn} \,. \end{align}
Thus, $K_{ij}$ is a geometric property of the hypersurface $\Sigma_{\T}$ and measures its curvature within the higher-dimensional manifold $\mathcal{M}$. Extrinsic curvature is meaningful only when the hypersurface is embedded in a larger space because it depends on the geometry of this higher-dimensional manifold via the covariant derivative of the normal vector. Geometrically, extrinsic curvature indicates how the hypersurface is curved within $\mathcal{M}$ by showing how the normal vectors at nearby points in $\mathcal{M}$ deviate from being parallel. Given the orthogonality condition $\ovec{e}_a \cdot \ovec{n} = 0$, the extrinsic curvature can also be expressed as
\begin{align} K_{ij}= \ovec{n} ~ \cdot ^4\nabla_i \ovec{e}_j = - X^{\mu}_i X^{\nu}_j n_{\mu;\nu} = - n_{i;j} \label{Kab2} \,. \end{align}
This expression emphasizes the role of the normal vector's derivative in defining the extrinsic curvature of the hypersurface within the larger manifold. Then the extrinsic curvature can be rewritten in terms of the lapse and shift functions as:
\begin{align}
K_{ij} &=-n_{i;j} =-n_{i,j}+^4\Gamma^{\mu}_{ij} n_{\mu} = ^4\Gamma^{0}_{ij} n_{0} = n_0 \left(g^{00}~^4\Gamma_{0ij}+g^{0c}~^4\Gamma_{cji} \right) \label{Kijgen} \,.
\end{align}
If we adopt $N= \tc/c$, $N_i = 0$, and Eq.~\eqref{ofn}, then we obtain
\begin{align}
&K_{ij} =- \frac{H}{\tc} g_{ij}  \quad , \quad 
K = -3 \frac{H}{\tc} \label{KRW} \,.
\end{align}
With the splittings and projections of the Riemann tensor associated with the manifold, we obtain results that will be useful when deriving the $3+1$ splittings of the Einstein tensor. 

\subsection{Gauss-Codazzi relation}
\label{subsec:GCrel}
%\textcolor{Red}{add Gauss-Codazzi relation using p.34 of Eric, Aug. 07} 
The Gauss and Codazzi relations are fundamental equations in differential geometry that link the intrinsic and extrinsic curvatures of a hypersurface embedded in a higher-dimensional manifold. We can obtain these relations by applying the Ricci identity to specific vectors associated with the hypersurface $\Sigma_{\T}$.

\begin{itemize}
\item Gauss relation:
The Gauss relation describes how the intrinsic curvature of the hypersurface $\Sigma_{\T}$ is related to the curvature of the ambient manifold and the extrinsic curvature of the hypersurface. It is derived by applying the Ricci identity to a tangent vector field $\vec{A}$ on $\Sigma_{\T}$ %\textcolor{red}{Eric 2.89}
\begin{equation}
\left( D_{\alpha} D_{\beta} - D_{\beta} D_{\alpha} \right) A^{\gamma} = R^{\gamma}_{~~\mu\alpha\beta} A^{\mu} = \left(K_{\alpha\mu} K^{\gamma}_{~\beta} - K_{\beta\mu} K^{\gamma}_{~\alpha} \right) A^{\mu} + p^{\rho}_{\alpha} p^{\sigma}_{\beta} p^{\gamma}_{\lambda} \,^{(4)} R^{\lambda}_{~\mu\rho\sigma} A^{\mu} \label{Gaussrel1} \,.
\end{equation}
From Eq.~\eqref{Gaussrel1}, we obtain the Gauss relation by components
\begin{align}
p^{\gamma}_{\rho} p^{\mu}_{\alpha} p^{\sigma}_{\delta} p^{\nu}_{\beta} \,^{(4)} R^{\rho}_{~~\mu\sigma\nu} = R^{\gamma}_{~~\alpha\delta\beta} + K^{\gamma}_{~\delta} K_{\alpha\beta} - K^{\gamma}_{~\beta} K_{\delta\alpha} \label{Gaussrel2} \,. 
\end{align}
We can also obtain the contracted Gauss relation by putting $\gamma = \delta$
\begin{align}
p^{\mu}_{\alpha} p^{\nu}_{\beta} \,^{(4)} R_{\mu\nu} + p_{\mu\alpha} p^{\nu}_{\beta} n^{\rho} n^{\sigma} \,^{(4)} R^{\mu}_{~~\rho\nu\sigma} = R_{\alpha\beta} + K K_{\alpha\beta} - K_{\alpha\delta} K^{\delta}_{~\beta} \label{conGaussrel1} \,. 
\end{align}
We multiply $p^{\alpha\beta}$ in Eq.~\eqref{conGaussrel1} to obtain the scalar Gauss relation
\begin{align}
\,^{(4)} R + 2 n^{\mu} n^{\nu} \,^{(4)} R_{\mu\nu} = R + K^2 - K_{ij} K^{ij} \label{conGaussrel2} \,. 
\end{align}
If we use equations \eqref{nalpha2},~\eqref{tRmp2}, \eqref{3DtRmp},  and \eqref{KRW},  then the left hand side and right hand side of the Eq.~\eqref{conGaussrel2} become
\begin{align}
&\,^{(4)} R + 2 n^{\mu} n^{\nu} \,^{(4)} R_{\mu\nu} = \frac{6}{\tc^2} \left( \frac{\ddot{a}}{a} + \frac{\dot{a}^2}{a^2} + k \frac{\tc^2}{a^2} - H^2 \frac{d \ln \tc}{d \ln a} \right) + 2 n^{0} n^{0} \,^{(4)} R_{00} + 2 n^{i} n^{i} \,^{(4)} R_{ij} \nonumber \\
%&=\frac{6}{\tc^2} \left( \frac{\ddot{a}}{a} + \frac{\dot{a}^2}{a^2} + k \frac{\tc^2}{a^2} - H^2 \frac{d \ln \tc}{d \ln a} \right) + 2 \cdot 1^2  \,^{(4)} R_{00} + 2 \cdot 0^2 \,^{(4)} R_{ij} \nonumber \\
&=\frac{6}{\tc^2} \left( \frac{\ddot{a}}{a} + \frac{\dot{a}^2}{a^2} + k \frac{\tc^2}{a^2} - H^2 \frac{d \ln \tc}{d \ln a} \right) - 6 \frac{6}{\tc^2} \left( \frac{\ddot{a}}{a} - H^2 \frac{d \ln \tc}{d \ln a} \right) \nonumber \\
&=\frac{6}{\tc^2} \left( \frac{\dot{a}^2}{a^2} + k \frac{\tc^2}{a^2} \right) \label{LHSGrel} \,, \\
&R + K^2 - K_{ij} K^{ij} = 6 \frac{k}{a^2} + \left( -3 \frac{H}{\tc} \right)^2 - \left(- \frac{H}{\tc} g_{ij} \right) \left(- \frac{H}{\tc} g^{ij} \right) \nonumber \\
&=\frac{6}{\tc^2} \left( \frac{\dot{a}^2}{a^2} + k \frac{\tc^2}{a^2} \right) \label{RHSGrel} \,.
\end{align}

\item Codazzi relation:
The Codazzi relation describes how the extrinsic curvature of the hypersurface is related to the ambient manifold's geometry. We can derive it by applying the Ricci identity to the normal vector $\ovec{n}$ on the $\Sigma_{\T}$. It links the derivatives of the extrinsic curvature to the ambient manifold's geometry. It is given by projecting the four-dimensional Riemann tensor acting on the normal vector into $\Sigma_{\T}$
\begin{align}
& p^{\mu}_{\alpha} \, p^{\nu}_{\beta} \, p^{\gamma}_{\rho} \left( \nabla_{\mu} \nabla_{\nu} - \nabla_{\nu} \nabla_{\mu} \right) n^{\rho} = p^{\mu}_{\alpha} \, p^{\nu}_{\beta} \, p^{\gamma}_{\rho} \,^{(4)} R^{\rho}_{~\sigma\mu\nu} n^{\sigma} = D_{\beta} K^{\gamma}_{\alpha} - D_{\alpha} K^{\gamma}_{\beta } \label{Codazzirel1} \,.
\end{align}
The contracted Codazzi relation is obtained from Eq.~\eqref{Codazzirel1} by contracting on the indices $\alpha$ and $\gamma$
\begin{align}
p^{\nu}_{\beta} n^{\sigma} \,^{(4)} R_{~\sigma\nu}  = D_{\beta} K - D_{\alpha} K^{\alpha}_{\beta } \label{Codazzirel2} \,.
\end{align}
\end{itemize}

\section{Einstein field equations in 3+1 formalism}
\label{sec:EEs}

In the $3+1$ formalism, EFEs are split into a set of constraint equations and evolution equations that govern the dynamics of spacetime. When applied to the RW metric, which describes a homogeneous and isotropic universe, these equations simplify significantly due to the symmetry of the metric. The momentum constraint typically vanishes in the RW metric due to the absence of preferred directions in space (\textit{i.e.}, $N^i = 0$). This holds for both the SMC and the meVSL model. The Hamiltonian constraint and evolution equations of $3$D spatial geometry show the effect of the VSL. 

\subsection{The Einstein equations}
\label{subsec:EEs}

We consider a spacetime $(\mathcal M\,,\tf{g})$ such that $\tf{g}$ obeys the Einstein equation 
\begin{align}
&{}^4 \tf{R} - \frac{1}{2} {}^4\!R \, \tf{g} = \kappa \tf{T} \quad \Rightarrow \quad {}^4 R_{\mu\nu} - \frac{1}{2} {}^4\!R g_{\mu\nu} = \kappa T_{\mu\nu} \label{dec:Einstein}  \,,
\end{align}
where ${}^4\!\tf{R}$ is the Ricci tensor associated with $\tf{g}$, ${}^4\!R$ is the corresponding Ricci scalar, and $\tf{T}$ is the matter stress-energy tensor.  We shall also use the equivalent form
\begin{align}
&{}^4\!\tf{R} = \kappa \left( \tf{T} - \frac{1}{2} T \, \tf{g} \right) \quad \Rightarrow \quad {}^4\! R_{\mu\nu} = \kappa \left( T_{\mu\nu} - \frac{1}{2} T \, g_{\mu\nu} \right) \label{dec:Einstein2} \,.
\end{align}
Let us assume that the spacetime $(\mathcal M\,,\tf{g})$ is globally hyperbolic and let be $\Sigma_{\mathcal T}$ by
a foliation of $\M$ by a family of spacelike hypersurfaces. 
The foundation of the 3+1 formalism amounts to projecting the Einstein equation~\eqref{dec:Einstein} onto $\Sigma_{\mathcal T}$ and perpendicularly to it. To this purpose let us first consider the 3+1 decomposition of the stress-energy tensor. 

\subsection{3+1 decomposition of the stress-energy tensor} 
\label{s:dec:T3p1}

The components of the stress-energy tensor of a perfect fluid $T_{\alpha\beta}$ is given by
\begin{align}
T_{\alpha\beta} &= \left( \rho + \frac{P}{\tc^2} \right) U_{\alpha} U_{\beta} + P g_{\alpha\beta} \quad , \quad U^{\alpha} = \frac{d x^{\alpha}}{d \tau} \quad , \quad U^{\alpha} = \tc n^{\alpha} \nonumber \\
&\equiv \left( \rho \tc^2 + P \right) n_{\alpha} n_{\beta} + P \left( \gamma_{\alpha\beta} - n_{\alpha} n_{\beta} \right) = \rho \tc^2 n_{\alpha} n_{\beta} + P \gamma_{\alpha\beta}  \label{Tp} \,.
\end{align}
From the definition of a stress-energy tensor, $\tf{T}$, we can decompose the components of $\tf{T}$ as
\begin{align}
&\vec{g} \, \tf{T} \equiv T_{\alpha\beta} = \delta^{\mu}_{~\alpha} \delta^{\nu}_{~\beta} T_{\mu\nu} = \left( p^{\mu}_{~\alpha} - n^{\mu}n_{\alpha} \right) \left( p^{\nu}_{~\beta} - n^{\nu}n_{\beta} \right) T_{\mu\nu}  \nonumber \\ 
&= p^{\mu}_{~\alpha} p^{\nu}_{~\beta} T_{\mu\nu} - \left( p^{\mu}_{~\alpha} n^{\nu} T_{\mu\nu} \right) n_{\beta} - \left( p^{\nu}_{~\beta}  n^{\mu} T_{\mu\nu} \right) n_{\alpha} + n_{\alpha} n_{\beta} n^{\mu} n^{\nu}T_{\mu\nu} \nonumber \\
&\equiv S_{\alpha\beta} + q_{\alpha} \, n_{\beta} + q_{\beta} \, n_{\alpha} + \rho \tc^2 n_{\alpha} n_{\beta}  \label{gT1}  \\
&= \left( p^{\mu}_{~\alpha} p^{\nu}_{~\beta} T_{\mu\nu} -\frac{1}{3} \gamma_{\alpha\beta} T \right) + \frac{1}{3} \gamma_{\alpha\beta} T   - \left( p^{\mu}_{~\alpha} n^{\nu} T_{\mu\nu} \right) n_{\beta} - \left( p^{\nu}_{~\beta}  n^{\mu} T_{\mu\nu} \right) n_{\alpha} + n_{\alpha} n_{\beta} n^{\mu} n^{\nu}T_{\mu\nu} \nonumber \\
&\equiv \pi_{\alpha\beta} + P \gamma_{\alpha\beta} + q_{\alpha} \, n_{\beta} + q_{\beta} \, n_{\alpha} + \rho \tc^2 \,n_{\alpha} n_{\beta} \label{gT2} \,. %\\ 
%&\Rightarrow \tf{T} \left( \vec{\gamma}\,,\vec{\gamma} \right) -  \vec{T} \left( \ovec{n} \,,\vec{\gamma} \right) \of{n} - \vec{T} \left( \ovec{n} \,,\vec{\gamma} \right) \of{n} + \tf{T} \left( \ovec{n}\,,\ovec{n}\right) \of{n} \otimes \of{n}  \nonumber \,.
\end{align}
Thus, $\tf{T}$ becomes
\begin{align}
\tf{T} &\equiv \tf{S} + \of{n} \otimes \of{q} + \of{q} \otimes \of{n} + \rho \tc^2 \,\of{n} \otimes \of{n} \label{tfT1} \\
&\equiv \tf{\pi} + P \tf{\gamma} + \of{n} \otimes \of{q} + \of{q} \otimes \of{n} + \rho \tc^2 \,\of{n} \otimes \of{n} \quad , \quad \tf{g} = \tf{\gamma} - \of{n} \otimes \of{n}  \label{tfT2} \\
&\rho c^2 \equiv \tf{T} \left( \ovec{n}\,,\ovec{n} \right) \quad \Rightarrow \quad \rho \tc^2 \equiv T_{\mu\nu} n^{\mu} n^{\nu} \,, \nonumber
\end{align}
where $\rho$ is the local mass density, $P$ is the isotropic pressure in the rest frame of $n^{\alpha}$, $\of{q}$ is the spatial part of the energy-momentum flux vector, and $\tf{\pi}$ is the anisotropic stresses. Trace of $\tf{T}$ is given by
\begin{align}
T &\equiv T^{\alpha}_{\alpha} = S^{\alpha}_{\alpha} + 2 n^{\alpha} q_{\alpha} + \rho \tc^2 n^{\alpha} n_{\alpha} = S + 2 \underbrace{\langle \of{n} \,, \ovec{q} \rangle}_{=0} +  \underbrace{\langle \of{n} \,, \ovec{n} \rangle}_{=-1} \rho \tc^2 = S - \rho \tc^2 \label{Ttrac} \\
&\equiv \pi^{\alpha}_{\alpha} + P \gamma^{\alpha}_{\alpha} + 2 n^{\alpha} q_{\alpha} + \rho \tc^2 n^{\alpha} n_{\alpha} = 0 + 3 P + 0 - \rho \tc^2 = 3P - \rho \tc^2 \nonumber \,.
\end{align}

\subsection{Projection of the Einstein equation} 
\label{s:dec:project_Einstein}

We are ready to project the Einstein equation \eqref{dec:Einstein} onto the hypersurface $\Sigma_{\mathcal{T}}$ and along its normal vector thanks to the $3+1$ decomposition of the stress-energy tensor and the equivalent decompositions of the spacetime Ricci tensor. This approach allows us to analyze the Einstein equations in terms of the geometry of the hypersurface and its evolution over time. When projecting the equations, we consider the following three possibilities:

\begin{itemize}
	\item Projection along the normal: This involves examining the components of the Einstein equations that are normal to the hypersurface. This projection explains how the hypersurface evolves in time, capturing the extrinsic curvature and its relation to the matter distribution.
	\item Mixed projections: This involves projecting onto the hypersurface and along its normal, capturing interactions between the geometry of the hypersurface and its embedding in the larger spacetime.
	\item Projection onto the hypersurface: This involves decomposing the Einstein equations into parts lying entirely within the hypersurface $\Sigma_{\mathcal{T}}$. This projection provides constraints related to the intrinsic curvature and matter content on the hypersurface.
\end{itemize}

By analyzing these projections, we can understand how the gravitational field influences the dynamics and geometry of the hypersurface, providing insight into the system's evolution in a $3+1$ framework.

\subsubsection{Full projection perpendicular to $\Sigma_{\T}$}
\label{subsub:FPpSigma}

It amounts to applying the Einstein equation~\eqref{dec:Einstein}, which is an identity between bilinear forms, to the couple $(\w{n},\w{n})$; we get, since $\w{g}(\w{n},\w{n})=-1$,

\begin{align}
&{}^4 \tf{R} - \frac{1}{2} {}^4\!R \, \tf{g} = \kappa \tf{T} \quad \Rightarrow \quad {}^4 R_{\mu\nu} - \frac{1}{2} {}^4\!R g_{\mu\nu} = \kappa T_{\mu\nu} \label{dec:Einstein-2}  \\
&{}^4 \tf{R} \left( \ovec{n}\,, \ovec{n} \right) - \frac{1}{2} {}^4\!R \, \tf{g} \left( \ovec{n}\,, \ovec{n} \right) = \kappa \tf{T} \left( \ovec{n}\,, \ovec{n} \right) \quad ,  \textrm{where} \quad g_{\mu\nu} n^{\mu} n^{\nu} = n_{\mu} n^{\mu} = -1 \quad\Rightarrow \nonumber \\ 
&{}^4 R_{\mu\nu} n^{\mu} n^{\nu} + \frac{1}{2} {}^4\!R = \kappa T_{\mu\nu} n^{\mu} n^{\nu} = \kappa \rho \tc^2 \nonumber \\
&R + K^2 - K_{ij} K^{ij} = 2 \kappa \rho \tc^2 \nonumber \,.
\end{align}
In the last equality, we use scalar Gauss relation
\begin{align}
2 {}^4 R_{\mu\nu} n^{\mu} n^{\nu} + {}^4\!R = R + K^2 - K_{ij} K^{ij} \label{sGauss} \,.
\end{align}
This equation relates the intrinsic curvature of the spatial hypersurface to the extrinsic curvature and matter content. It is called the Hamiltonian (Gauss) constraint, a scalar one. We can explicitly write this equation for the RW metric as 
\begin{align}
&\frac{6}{\tc^2} \left( \frac{\dot{a}^2}{a^2} + k \frac{\tc^2}{a^2} \right) = \frac{16 \pi G}{\tc^4} \rho \tc^2 \quad \Rightarrow  \quad \frac{\dot{a}^2}{a^2} + k \frac{\tc^2}{a^2} = \frac{8 \pi G}{3} \rho \label{PSEFE} \,.
\end{align}
This result is, of course, identical to the Einstein equations in the meVSL model \cite{Lee:2020zts,Lee:2022heb,Lee:2023bjz,Lee:2024part}.

\subsubsection{Mixed projection}
\label{subsub:MixedPro}

Let us project the Einstein equation (\ref{dec:Einstein}) once onto $\Sigma_{\T}$ and once along the normal $\ovec{n}$:
\begin{align}
&{}^{(4)} \tf{R} - \frac{1}{2} {}^{(4)}\!R \, \tf{g} = \kappa \tf{T} \quad \Rightarrow \quad {}^{(4)} R_{\mu\nu} - \frac{1}{2} {}^{(4)}\!R g_{\mu\nu} = \kappa T_{\mu\nu} \label{dec:Einstein-3}  \\
&{}^{(4)} \tf{R} \left( \ovec{n}\,, \vec{p} \right) - \frac{1}{2} {}^{(4)}\!R \, \tf{g} \left( \ovec{n}\,, \vec{p} \right) = \kappa \tf{T} \left( \ovec{n}\,, \vec{p} \right) \quad ,  \textrm{where} \quad p^{\mu}_{~\alpha} n^{\nu} g_{\mu\nu} = p^{\mu}_{~\alpha} n_{\mu} = 0 \quad \Rightarrow \nonumber \\ 
&p^{\mu}_{~\alpha} n^{\nu} {}^{(4)} R_{\mu\nu} + \frac{1}{2} {}^{(4)}\!R \gamma^{\mu}_{\alpha} n^{\nu} g_{\mu\nu}  = \gamma^{\mu}_{\alpha} n^{\nu} {}^{(4)} R_{\mu\nu} = \kappa \gamma^{\mu}_{\alpha} n^{\nu} T_{\mu\nu} = - \kappa q_{\alpha} \nonumber \\
&D_{\mu} K^{\mu}_{\alpha} - D_{\alpha} K = \kappa q_{\alpha} \nonumber
\end{align}
In the last equality, we use the contracted Codazzi equation
\begin{align}
p^{\mu}_{~\alpha} n^{\nu} {}^{(4)} R_{\mu\nu} = D_{\alpha} K - D_{\mu} K^{\mu}_{~\alpha} \label{cCodazzi}.
\end{align}
This is called the momentum (Codazzi) constraint.  This equation ensures the consistency of the embedding of the spatial hypersurface in spacetime. No equation is obtained from this equation for the background RW metric. 

\subsubsection{Full projection onto $\Sigma_{\T}$}
\label{subsub:FPoSigma}

This amounts to applying the operator $\vec{p}^{\ast}$ to the Einstein equation.
\begin{align}
&\vec{p}^{\ast} {}^{(4)}\!\tf{R} = \kappa \left( \vec{p}^{\ast} \tf{T} - \frac{1}{2} T \, \vec{p}^{\ast} \tf{g} \right) \quad \textrm{where} \quad S_{\alpha\beta} = \pi_{\alpha\beta} + P \gamma_{\alpha\beta} \Rightarrow \nonumber \\ 
&p^{\mu}_{~\alpha} p^{\nu}_{~\beta} {}^{(4)}\! R_{\mu\nu} = \kappa \left( p^{\mu}_{~\alpha} p^{\nu}_{~\beta}  T_{\mu\nu} - \frac{1}{2} T \, p^{\mu}_{~\alpha} p^{\nu}_{~\beta} g_{\mu\nu} \right) = \kappa \left( S_{\alpha\beta} - \frac{1}{2} T \, \gamma_{\alpha\beta}  \right)  \label{dec:Einstein2-2} , \\
& \quad = -\frac{1}{N} \mathcal{L}_{m} K_{\alpha\beta} - \frac{1}{N} D_{\alpha\beta} + ^{3} R_{\alpha\beta} + K K_{\alpha\beta} - 2 K_{\alpha\mu} K^{\mu}_{~\beta} \nonumber 
\end{align}
where we use
\begin{align}
 \frac{1}{N} \mathcal{L}_{m} K_{\alpha\beta} = p^{\mu}_{~\alpha} p^{\nu}_{~\beta} n^{\sigma} \nabla_{\sigma} K_{\mu\nu} - 2 K_{\alpha\mu} K^{\mu}_{~\beta} \label{LiemK} 
\end{align}
It is called an evolution equation. Equation~\eqref{dec:Einstein2-2} is a rank-$2$ tensorial (bilinear forms) equation within $\Sigma_{\T}$, with six independent components. The above equation for the RW metric becomes
\begin{align}
&p^{\mu}_{~\alpha} p^{\nu}_{~\beta} {}^{(4)}\! R_{\mu\nu} = \kappa \left( S_{\alpha\beta} - \frac{1}{2} T \, \gamma_{\alpha\beta}  \right) \quad \Rightarrow \quad p^{i}_{~j} p^{i}_{~j} {}^{(4)}\! R_{ij} = \kappa \left( S_{ij} - \frac{1}{2} T \, \gamma_{ij}  \right) \nonumber \\
&\frac{g_{ij}}{\tc^2} \left( 2 \frac{\dot{a}^2}{a^2} + \frac{\ddot{a}}{a} + 2 k \frac{\tc^2}{a^2} - H^2 \frac{d \ln \tc}{d \ln a} \right) = \frac{4 \pi G}{\tc^4} \left( 2 P - 3P + \rho \tc^2   \right) g_{ij} \nonumber \\ 
&\left( 2 \frac{\dot{a}^2}{a^2} + \frac{\ddot{a}}{a} + 2 k \frac{\tc^2}{a^2} - H^2 \frac{d \ln \tc}{d \ln a} \right) = \frac{4 \pi G}{\tc^2} \left(  \rho \tc^2 - P \right) \label{FPEFE-2} \,.
\end{align}
where we use $p^0_{~0} = 0$\,, $p^{i}_{~j} = \delta^{i}_{~j}$, and $p^{0}_{~j} = 0$.  
If we combine Eqs.~\eqref{PSEFE} and~\eqref{FPEFE-2}, then we obtain
\begin{align}
\frac{\ddot{a}}{a} = -\frac{4 \pi G}{3} \left( \rho + \frac{P}{\tc^2} \right) + H^2 \frac{d \ln \tc}{d \ln a} \label{ddotaoa} \,. 
\end{align}
Again, this is dentical to the Einstein equation for the acceleration of the universe in the meVSL model \cite{Lee:2020zts,Lee:2022heb,Lee:2023bjz,Lee:2024part}.

\section{Discussion and Conclusions}
\label{sec:con}

The $3+1$ formalism offers several advantages over directly solving the Einstein equations in the Robertson-Walker metric within the standard cosmology model. It provides a clear separation of space and time, facilitating the interpretation of physical quantities. It also offers flexibility in choosing coordinates and gauges and aids in the physical interpretation of observers. It simplifies the analysis of cosmological perturbations, making it a powerful tool for both theoretical and practical applications in cosmology.

These characteristics also apply equally well to the minimally extended varying speed of light (meVSL) model. The shift vector is zero In both the standard model of cosmology and the meVSL model for the Robertson-Walker metric. $N^i = 0$ is essential for maintaining the homogeneity and isotropy of the Robertson-Walker metric. However, while the lapse function is $1$ in the standard model of cosmology, it becomes a function of time in the meVSL model, expressed as $N = \tilde{c}/c = 1 + b/4 Ht$. In the $3+1$ formalism, we can interpret it as a change in the speed of light on the hypersurfaces. Meanwhile, we have understood it as cosmological time dilation described by $1/a^{1-b/4}$ in Einstein equations. We plan to apply the results of the background evolution in the meVSL model obtained using the $3+1$ formalism to perturbation calculations in our forthcoming manuscript.

\section{Acknowledgments}
This research was funded by the National Research Foundation of Korea (NRF), funded both by the Ministry of Science, ICT, and~Future Planning (Grant No. NRF-2019R1A6A1A10073079) and by the Ministry of Education (Grant No. NRF-RS202300243411).

\section{Refrences}

\end{document}